\documentclass[pageno]{jpaper}
\usepackage{listings, color, setspace, graphics}
\usepackage{fancyhdr, amsmath, amssymb, setspace, amssymb}
\usepackage[10pt]{moresize}
\usepackage{graphicx}
\usepackage{balance}  
\usepackage{amsmath}
\usepackage{mathptmx}
\usepackage{nicefrac}
\usepackage{units}
\usepackage{float}
\usepackage{textcomp}
\usepackage{verbatim}
\usepackage{pstricks}
\usepackage{booktabs}
\usepackage{times}
\usepackage{epsfig}
\usepackage{multirow}
\usepackage{mdwlist}
\usepackage{hyperref}
\usepackage{amssymb}
\usepackage{balance}
\usepackage[T1]{fontenc}
\usepackage{algorithm}
\usepackage{algorithmic} 
\usepackage{enumitem}
\usepackage{setspace}
\usepackage{calc}
\usepackage{euscript}
\usepackage{xcolor}
\usepackage{caption}
\usepackage{xspace}
\usepackage{array}
\usepackage[sort,nocompress]{cite}
\usepackage{wrapfig,lipsum,booktabs}

\renewcommand{\algorithmiccomment}[1]{\bgroup\hfill//~#1\egroup}

\usepackage[scaled=0.85]{beramono}

\usepackage[normalem]{ulem}
\usepackage{authblk}

\newcommand{\projectname}{{Hop}\xspace}

\begin{document}

\title{\projectname:  Heterogeneity-aware Decentralized Training}
\author[1]{Qinyi Luo}
\author[2]{Jinkun Lin}
\author[1]{Youwei Zhuo}
\author[1]{Xuehai Qian}
\affil[1]{University of Southern California, CA, USA}
\affil[2]{Tsinghua University, Beijing, China}

\date{}
\maketitle

\thispagestyle{empty}

\begin{abstract}
Recent work has shown that decentralized algorithms can deliver superior performance over centralized ones in the context of machine learning. The two approaches, with the main difference residing in their distinct communication patterns, are both susceptible to performance degradation in heterogeneous environments. Although vigorous efforts have been devoted to supporting centralized algorithms against {\em heterogeneity}, little has been explored in decentralized algorithms regarding this problem. 

This paper proposes {\em \projectname}, the first heterogeneity-aware decentralized training protocol.
Based on a unique characteristic of decentralized training that we have identified, the {\em iteration gap}, we propose a queue-based synchronization mechanism that can efficiently implement backup workers and 
bounded staleness in the decentralized setting. 
To cope with
deterministic slowdown, we propose skipping iterations 
so that the effect of slower workers is further mitigated.
We build a prototype implementation of 
\projectname on \textsc{TensorFlow}.
The experiment results on CNN and SVM
show significant speedup over 
standard decentralized training in heterogeneous settings.

\end{abstract}

\section{Introduction}
\label{sec:intro}

Machine Learning (ML) has proven to be an effective way to automatically extract information from data. Great success has been achieved in various domains including speech recognition \cite{Hinton:speech-recognition}, image classification \cite{CVPR15:image}, natural language processing \cite{JMLR11:NLP}, etc. Obtaining useful ML models requires training with large data sets for long hours. With the increase of model and data size, 
distributed training is the current solution to achieve 
speedups with more compute nodes. 

The most commonly used algorithm in distributed training is Stochastic Gradient Descent (SGD). It is an iterative algorithm that computes gradients over a small fraction of data in every iteration and modifies model parameters accordingly until convergence. 
Two approaches exist to execute it in distributed
fashion --- centralized and decentralized.

In the centralized approach, central nodes called Parameter Servers (PS) are responsible for maintaining the parameters, while other machines, called workers, are responsible for computation. Current available protocols mainly include Bulk Synchronous Parallel (BSP) \cite{GraphX:BSP,NSDI12:BSP,AAAI15:BSP}, Stale Synchronous Parallel (SSP) \cite{NIPS2013_SSP,Petuum:SSP,SIGMOD16:SSP} and asynchronous coordination \cite{NIPS2011_hogwild,OSDI14:async,NIPS2012_async}. 
In the decentralized approach, every machine maintains its own version of parameters and coordinates according to a specified communication graph, which can be dense, as in All-Reduce \cite{DBLP:allReduce,OSDI08:allReduce,PPoPP03:allReduce}, or sparse \cite{ArXiv_ASAP,NIPS2017_dPSGD,ArXiv_ADPSGD,DBLP:conf/nips/send_difference}.

Performance bottleneck in distributed training largely comes from two sources: 
{\em a)} communication hotspot, especially for the PS --- all workers need to put/get updates
to/from the PS; and 
{\em b)} heterogeneity, which causes fast workers to wait for slow ones during synchronization. The first problem can be alleviated with decentralized training: since the amount of communication at a node is determined by the degree of a graph, the PS is no longer the bottleneck. As a result, decentralized training has recently shown superior performance over centralized training \cite{NIPS2017_dPSGD}. However, it falls short on support for heterogeneity, although various approaches have been proposed in centralized training to deal with this problem \cite{NIPS2011_hogwild,NSDI2017_Gaia,SIGMOD2017_het,NIPS2013_SSP,ICLR2016_backup_workers}. Moreover, compared to centralized training, decentralized training receives much less support from ML systems like TensorFlow \cite{DBLP:Tensorflow}, MXNet \cite{CoRR15:MXNet} and CNTK \cite{KDD16:CNTK}, which makes it difficult to apply this less explored but potentially superior approach to real world problems.

To fill this gap, this paper proposes {\em \projectname}, a {\em heterogeneity-aware} decentralized training protocol.
We start with identifying a unique characteristic of decentralized training, the {\em iteration gap}, and determine its upper-bound based on communication graph among workers.
This property affects the cost and correctness
of an implementation.
We show that one of the current designs, 
\textsc{notify-ack}~\cite{ArXiv_ASAP}, in 
fact conservatively bounds the gap, but sacrifices the performance and implementation flexibility. 
We propose a queue-based synchronization mechanism for distributed coordination. Decentralized training 
can be realized using update queues sized 
based on the iteration gap, or using both update
and token queues which can control the iteration gap.
In essence, our general approach covers \textsc{notify-ack}
as a special case.

To mitigate the effects of heterogeneity, 
we propose backup workers and bounded staleness
in the decentralized setting.
Importantly, we show that 
the queue-based synchronization is {\em indispensable} in realizing these concepts
due to fundamental reasons. 
With backup workers, the iteration gap can 
become arbitrarily large; therefore, the 
controlled gap is no longer an optimization but
an necessity. 
For bounded staleness, queue-based synchronization
allows flexible iteration gap between workers
so that the benefits of staleness can be achieved. 
To cope with 
deterministic slowdown, we propose skipping iterations 
to further limit the effect of individual slow workers.
With the above key contributions, 
\projectname significantly advances the 
state-of-the-art of distributed training. 

We build a prototype implementation of 
\projectname on \textsc{TensorFlow}. 
We evaluate the system on CNN and SVM applications. Our results show that our system achieves better performance
than centralized training with PS.
In case of random and deterministic slowdown, our proposed features, backup workers, bounded staleness and skipping iterations, achieved significant speedups over standard decentralized training. In addition, 
we have observed an interesting result that carefully designed communication graphs with a smaller spectral gap can 
even perform better in heterogeneous network settings.

\section{Background and motivation}
\label{sec:back}
The problem considered in this paper is to use SGD to minimize a loss function $F$ over a data set $S$, and the update function in each iteration is given by
$x \leftarrow x - \eta \cdot \nabla_x F(x;\xi)$,
where $\xi$ is a mini-batch of data randomly sampled from $S$ and model parameters are represented by $x$. 
this process can be parallelized to be executed in a distributed 
environment~\cite{Zinkevich2010PSGD}. A well-known mechanism for parallel SGD is training with Parameter Servers (PS) \cite{Li2014ParamServer}.

\subsection{Distributed Training with Parameter Server}
\label{sec:trainPS}
Training with PS involves choosing one or a few central nodes as PS that are responsible for maintaining and updating model parameters \cite{Li2014ParamServer}. Other machines, called workers, pull parameters from the PS, compute gradients based on random samples and send gradients back to the PS. Then the PS will update the parameters based on the received gradients.

In the most basic setting, workers are synchronized at the end of each iteration. They are not allowed to pull new parameters from the PS until the PS has received updates from every worker and applied them to the parameters. In this way, workers always work on the same and most up-to-date version of parameters. However, there are two main drawbacks of the synchronous setting: {\em a)} fast workers always have to wait for slow ones, which is called the straggler problem; and {\em b)} communication bottlenecks or hotspots easily occur at the PS.

The current approach to mitigate the communication bottleneck is 
to apply ring All-Reduce~\cite{ring_allreduce1,Ring_allreduce2} among all workers, without using 
PS. With the careful overlapping of communication and computation,
the communication hotspot at PS is eliminated. 
Logically, it implements All-Reduce, 
which means that the update
from one worker is {\em broadcast} 
to all other workers by the 
end of the iteration. 
While ring All-Reduce hides some communication latency with
computation, the actual latency can potentially be
increased when 
the single update from each worker travels in the ring and reaches all other workers.

\subsection{Decentralized Training}

It has been recently theoretically shown for the first time that decentralized algorithms can outperform centralized ones \cite{NIPS2017_dPSGD}. The two types of algorithms share the same order of computational complexity \cite{NIPS2017_dPSGD}, but decentralized algorithms enjoy faster communications since the communication load is spread across the graph instead of concentrated at the PS.

Although decentralized algorithms were studied before, its advantage was long hidden. Prior work \cite{2012_dualAveraging} showed that as the number of workers increases, it takes more iterations to reach a certain accuracy, based on the assumption that $F$ is convex. However, in the recent work \cite{NIPS2017_dPSGD}, convexity of $F$ was not assumed and it was shown that the convergence rate exhibits an asymptotically linear speedup with respect to the number of workers. The improved result serves the motivation 
for investigating decentralized training.

In decentralized training algorithms \cite{NIPS2017_dPSGD,ArXiv_ASAP,DBLP:conf/nips/send_difference}, there is no central node; every worker maintains its own version of parameters. Workers communicate with one another based on a predefined network topology. In each iteration, a worker computes gradients, sends its parameters to its out-going neighbors, and updates its parameters by averaging them with its in-coming neighbors. It remains a choice whether the gradients are applied to the parameters before or after the parameters are sent. The algorithm is shown in Figure \ref{alg0}. 
In the algorithm, parameters are sent before applying the gradients, which enables parallel execution of step 1 and 2 (the parallel approach). An alternative approach is to swap step 3 and 4, so that parameters are sent after applying the gradients (the sequential approach).
We will refer to this algorithm as standard decentralized training, and will discuss the two variants in Section~\ref{sec:framework}.

\begin{figure}
\centering
\small
 \begin{algorithmic}[1]
 \REQUIRE A set of worker nodes $V$ and their connection represented in a weighted adjacency matrix $W$
 \FOR{worker $i \in V$}
 \STATE Compute gradients over randomly selected samples $g_{k,i} = \nabla F(x_{k,i};\xi_{k,i})$
 \STATE Average parameters with my neighbors $x_{k+\frac{1}{2},i} \leftarrow \sum_{j \in V} W_{ji}x_{k,j}$
 \STATE Apply gradients $x_{k+1,i} \leftarrow x_{k+\frac{1}{2},i}-\eta_k \cdot g_{k,i}$
 \ENDFOR
 \end{algorithmic}
\caption{Standard Decentralized Training. }
\label{alg0}
\end{figure}

\subsection{System Heterogeneity}

As discussed in Section~\ref{sec:trainPS}, a main source of performance degradation is the straggler problem, which is 
an example of system heterogeneity. 
In general, it involves random aspects such as slowdown caused by resource sharing and hardware faults, as well as deterministic factors including differences in hardware computational capabilities and network bandwidths \cite{NSDI2017_Gaia,SIGMOD2017_het,ArXiv_ADPSGD}.

Fundamentally, both PS and ring All-Reduce lack 
the {\em flexibility} to tackle the heterogeneous execution environment due to the {\em fixed} communication pattern between workers and PS or between workers themselves. 
For the PS setting, previous work has proposed several ways to deal with this problem, e.g., updating parameters asynchronously \cite{NIPS2011_hogwild}, using backup workers \cite{ICLR2016_backup_workers}, allowing bounded staleness \cite{NIPS2013_SSP}, dynamically adjusting the learning rate \cite{SIGMOD2017_het}, and sending accumulated gradients when they reach a significance threshold \cite{NSDI2017_Gaia}.
In ring All-Reduce, the more restrictive communication 
pattern makes it impossible to implement some 
techniques, e.g., backup workers. 
In fact, the execution may suffer more 
from slow communication links and/or stragglers in the ring.


For decentralized training, which has gained interests only recently, relatively few efforts have been devoted to improving performance in heterogeneous environments. A fairly recent work \cite{ArXiv_ADPSGD} proposed an asynchronous scheme where every worker averages parameters with a randomly selected neighbor instead of all the in-coming neighbors.
However, as will be discussed in 
detail in Section~\ref{sec:opt4het}, it may lead to deadlock and can only 
work for a specific type of communication graphs.

\subsection{Challenges and Motivation}

Decentralized algorithms can outperform centralized ones, because it eliminates the communication hotspot at the PS. However, heterogeneity remains a problem, since workers still need to synchronize with its neighbors, and thus the straggler effect exists. It can be imagined that the influence of one slow worker or network link can spread to the whole graph through connected nodes.

In the next section, we will analyze the nature of distributed synchronization, propose heterogeneity-aware algorithms based on
insights from solutions for PS, and implement a distributed training system based on the proposed mechanisms.  

\section{Distributed Coordination}
\label{sec:framework}

This section analyzes the computation graph and the
existing protocol, i.e., \textsc{notify-ack}~\cite{ArXiv_ASAP},
for distributed training.
Then, we explain and prove an important property 
about the iteration gap between workers. 
Finally, we propose backup workers and bounded staleness
in distributed training based on PS to mitigate the effects
of heterogeneity. 
We demonstrate that \textsc{notify-ack},
the existing scheme for 
decentralized training, cannot 
support them, motivating the proposed queue-based synchronization
(Section~\ref{sec:queue}).


\subsection{Notations}

We define the communication topology among workers in distributed 
training as a weighted directed graph \(G = (V, E)\), 
where each node represents a worker. 
At each node $i$, $N_{in}(i) = \{j|(j,i) \in E\}$ and
$N_{out}(i) = \{j|(i,j) \in E\}$ denote the 
set of in-coming and out-going neighbors, respectively. 
An edge \(e = (i,j) \in E\) indicates that worker $i$ needs to send updates to worker $j$ during training.
The weight of an edge reflects 
how much ``influence'' the updates have upon worker $j$.
Each worker maintains a copy of all the parameters and
the local update is always assumed to be available --- 
there is a self-loop at every node \cite{ArXiv_ASAP,NIPS2017_dPSGD}, i.e., for all $i \in V$, $(i,i) \in E$.

Let $u_i$ denote an update generated by worker $i$,
then the aggregated update at worker $j$ is given by $\sum_{i \in N_{in}(j)} W_{ij}u_i$, where
$W$ is the weighted adjacency matrix of $G$. 
Previous works \cite{NIPS2017_dPSGD, ArXiv_ASAP} show 
that for decentralized training to perform well, $G$ must be connected and $W$ has to be doubly stochastic, i.e., the row sums and column sums of $W$ must be one. Normally, every update has the same influence: 
\begin{equation}
    \label{eq:average}
    W_{ij} = \Big\{
    \begin{tabular}{lc}
    $1/|N_{in}(j)|$ & for $i \in N_{in}(j)$ \\
    0 & otherwise
    \end{tabular}
\end{equation}
An update sent from worker $i$ to worker $j$ is denoted by $u_{i \rightarrow j}$.
If the update is generated in the $k$-th iteration, we further indicate the time stamp with $u_{i \rightarrow j}(k)$. 

\subsection{Computation Graph in Decentralized Training}

This section discusses two variants of computation graphs at a worker for 
decentralized training and the trade-off. 
The computation in an iteration involves five operations. 

\begin{itemize}[leftmargin=*]
    \item {$Compute$:} The worker consumes a randomly selected batch of data and computes gradients based on its current model parameters.
    
    \item {$Send$:} The worker sends its current parameters to its out-going neighbors. This operation is non-blocking; the worker can send its updates regardless of the status of its out-going neighbors. 
    We use $Send(i)$ to specify the send operations performed  
    in the $i$-th iteration. 
    
    \item {$Recv$:} The worker receives the model parameters of 
    its in-coming neighbors. Note that the worker does not request
    the parameters; instead, they are sent by the in-coming neighbors 
    proactively.
    We use $Recv(i)$ to specify that the received parameters are sent 
    in the $i$-th iteration. 
    The $Recv$ operation blocks until the parameters are completely 
    received. 
    
    \item {$Reduce$:} The worker averages the parameters it has received with its own parameters.
    
    \item {$Apply$:} The worker applies gradients to its current parameters, either before or after the $Reduce$.
\end{itemize}


Next, we describe two computation graphs used in 
recent works~\cite{NIPS2017_dPSGD,ArXiv_ASAP,DBLP:conf/nips/send_difference}, which are consistent with the algorithm described in Figure \ref{alg0}.

\begin{figure}
    \centering
    \includegraphics[width=\linewidth]{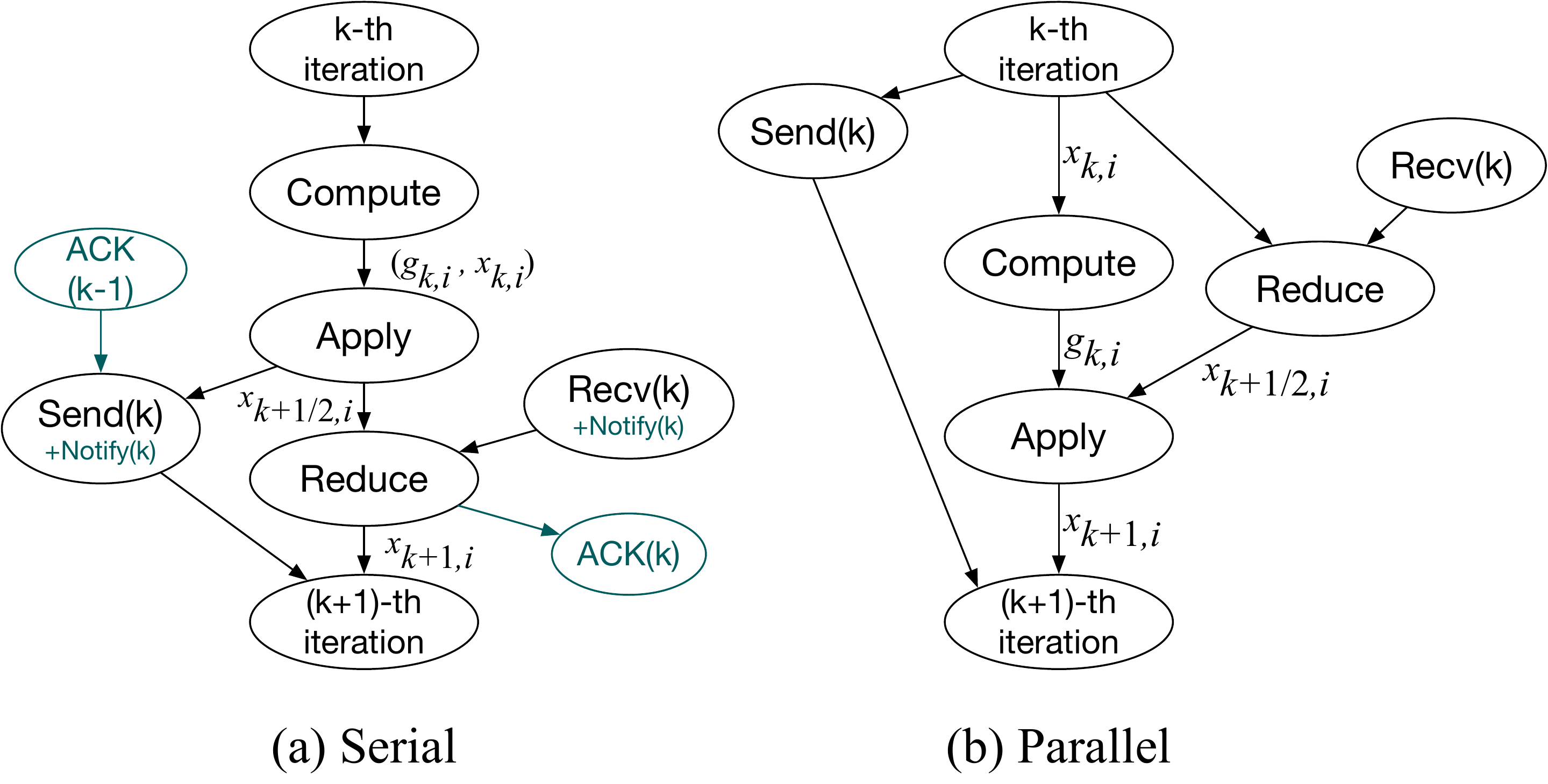}
    \caption{Computation Graph in Decentralized Training.}
    \label{fig:dec2ways}
\end{figure}

{\bf Serial Approach} 
Illustrated in Figure \ref{fig:dec2ways} (a),
upon entering a new iteration, each worker will $Compute$ gradients,
$Apply$ the gradients to its current parameters, and then $Send$ the new parameters to its out-going neighbors. When it has received new parameters from all its in-coming neighbors sent in the same iteration, it will perform a $Reduce$ and update the local parameters with the results.
This mode is adopted by \cite{ArXiv_ASAP}.

{\bf Parallel Approach} 
Shown in Figure \ref{fig:dec2ways} (b), each worker will $Send$ 
its current parameters at the beginning of an iteration, and at the same time $Compute$ gradients based on the same set of parameters. After receiving the parameters from its in-coming neighbors, it performs a $Reduce$, followed by an $Apply$, producing the local parameters for
update after applying the gradients to the reduced values.
This mode is used in \cite{NIPS2017_dPSGD,DBLP:conf/nips/send_difference}.


Compared to the serial approach, 
the parallel approach allows parallel execution of the $Compute$ and the $Reduce$.
However, the parallelism is achieved at the cost of 
{\em inaccurate gradients}.
Specifically, in Figure \ref{fig:dec2ways} (b),
the gradients are computed using the parameters {\em before} the $Reduce$
but are applied to the parameters {\em after} the $Reduce$.
This can harm the effectiveness of gradient descent.
On the contrary, the serial approach in Figure \ref{fig:dec2ways} (a) 
ensures that the gradients are generated with and applied to the {\em same} set of parameters. 
Therefore, the parallel approach enjoys faster iterations but requires more iterations to converge, while the serial approach needs fewer 
but longer iterations to converge. 
It reflects the interesting trade-off between the execution efficiency 
and statistic efficiency~\cite{Zhang:dinnwitted}.
We use parallel approach in our design.



\subsection{Iteration Gap and the Mixed-Version Problem}

An important characteristic of decentralized training is 
the potential large iteration gap, i.e., at any given time, workers can be found in a large range of iterations. In the centralized setting, no such gap exists in synchronous computation, because workers have to synchronize at the end of each iteration, which ensures that all workers always stay in the same iteration. For asynchronous computation, convergence is only guaranteed when bounded staleness is enforced, which sets a fixed upper-bound on the iteration gap between the fastest worker and the slowest one.

However, in decentralized setting, we show that the size of the gap is only limited by the graph topology and that large gaps can occur even in synchronous training. Before delving into the details, we first establish a basic and natural assumption.

\textbf{Assumption}. A worker can advance to the next iteration if and only if all of the following conditions are true: 
{\em a)} it has finished the computation of the current iteration; 
{\em b)} it has sent updates of the current iteration to its out-going neighbors; 
{\em c)} it has received updates required by the current iteration from its in-coming neighbors.

The above assumption was adopted in previous work \cite{ArXiv_ASAP}. Note that the assumption does {\em not} impose a global barrier between adjacent iterations. In fact, a global barrier can hurt the performance as it 
introduces unnecessary waiting: in order to enter the next iteration, a worker has to wait for all other workers to complete the current iteration, when actually it only needs to wait for its in-coming neighbors. Based on the assumption, an important result on iteration gap follows:

\textbf{Theorem 1}. Under the above assumption, at any given time, the maximum iteration difference between 
worker $i$'s iteration and worker $j$'s iteration is the length of the shortest path from node $j$ to node $i$, i.e.,
$Iter(i) - Iter(j) \leq length(Path_{j \rightarrow i})$,
where $Iter(i)$ is the iteration of worker $i$ for any $i \in V$ and  $length(Path_{j \rightarrow i})$ stands for the length of the shortest path from node $j$ to node $i$.

\textit{Proof}. The proof of the theorem is based on 
a simple observation: the maximal iteration difference between a node and its in-coming 
neighbor is 1, i.e., for any $i \in V$ and any $j \in N_{in}(i)$, $Iter(i) - Iter(j) \leq 1$. This is because worker $i$ can only advance to $Iter(i)$ when it has received worker $j$'s update of iteration $Iter(i)-1$. Note that we cannot derive the lower-bound of $Iter(i) - Iter(j)$ given only the directed edge from $j$ to $i$, because $Iter(j)$ can be much larger than $Iter(i)$ if worker $i$ is slower than worker $j$.

Now for two arbitrary nodes $i$ and $j$, consider the shortest path from $j$ to $i$. Going from $j$ to $i$, based on the observation above, every time we pass a node $v$, the maximal possible value of $Iter(v) - Iter(j)$ is increased by 1. Since $Iter(j) - Iter(j) = 0$ and there are $length(Path_{j \rightarrow i})$ other nodes on the path, we have $Iter(i) - Iter(j) \leq length(Path_{j \rightarrow i})$. $\blacksquare$

The existence of the iteration gap creates the mixed-version problem \cite{ArXiv_ASAP}, i.e., a worker can receive updates of various iterations at the same time from its in-coming neighbors. The problem may not be 
severe if the network is small, but as the size of the network grows, $length(Path_{i \rightarrow j})$ can be large, where $j \in N_{in}(i)$. In such cases, there are a large number of $u_{j \rightarrow i}$'s generated by worker $j$ but not consumed by worker $i$.

While this paper is the first to present and prove the theorem,
we find that a previous solution in ~\cite{ArXiv_ASAP} indeed provides 
a mechanism to bound the gap between workers~\footnote{Although ~\cite{NIPS2017_dPSGD} also implements decentralized training,
its focus is mainly on the algorithm, not implementation. We cannot find much detail to judge how
it handles the mix-version and iteration gap problem.
Thus, we only consider ~\cite{ArXiv_ASAP} regarding implementation.}. 
Specifically, a worker is prevented to send an update unless it has received confirmation that the previous update has been consumed. As illustrated in Figure \ref{fig:dec2ways} (a), this method, called \textsc{notify-ack}, requires a worker to send an $ACK$ message after the $Reduce$ to all its in-coming neighbors, announcing that the parameters they sent have been consumed. The in-coming neighbors will not perform the next $Send$ unless it has received the $ACK$. 
Essentially, in addition to the forward 
dependence from sender to receiver explained in 
{\em Theorem 1}, 
\textsc{notify-ack} also enforces the {\em backward}
dependence from receiver to sender. 
It leads to the over-restrictive iteration gap
between adjacent workers for the following reasons.


For $j \in N_{in}(i)$, we already know from \textit{Theorem 1} that $Iter(i) - Iter(j) \leq 1$. As for $Iter(j) - Iter(i)$, since worker $j$ needs to receive an ACK from worker $i$'s $(Iter(j)-1)$-th iteration in order to advance to iteration $Iter(j)+1$, the difference of their iterations is at most 2.

As for an arbitrary pair of workers $(i,j)$, the upper-bound on $Iter(i) - Iter(j)$ cannot be merely expressed by a function of $length(Path_{j \rightarrow i})$. 
This is because on any path between $i$ and $j$, we must ensure that $-2 \leq Iter(u) - Iter(v) \leq +1$ holds for any $v \in N_{in}(u)$ on the path. 
In another word, every time we pass a node $u$ 
from $v$, worker $u$ can be at most 1 iteration ahead of,
{\em and} at most 2 iterations behind worker $v$. 
As the result, the upper-bound of $Iter(i) - Iter(j)$,
i.e., how much worker $i$ is ahead of worker $j$,
is the {\em minimum} of maximum iteration gap following 
either $Path_{j \rightarrow i}$ or $Path_{i \rightarrow j}$ subject to the constraint between $u$ and $v$:
$Iter(i) - Iter(j) \leq min(length(Path_{j \rightarrow i}),2 \times length(Path_{i \rightarrow j}))$, --- more restrictive
than the iteration gap determined by {\em Theorem 1}.

Although \textsc{notify-ack} can ensure the sequential order of updates at the receiving worker, the tightly bounded iteration gap makes it an undesirable choice for heterogeneous environments where workers are expected to advance at various speeds. An intuitive example is that a fast worker may wait for a slow out-going neighbor's $ACK$ after it has received all the updates from its in-coming neighbors, 
--- in fact ready to advance to the next iteration.
To cope with heterogeneity, 
we propose two mechanisms, 
backup workers and bounded staleness, to accommodate 
larger iteration gaps.
While similar mechanisms exist in centralized training~\cite{ICLR2016_backup_workers,NIPS2013_SSP},
they have not been applied in the decentralized setting.
As we will show, both the protocol and implementation will need to be carefully redesigned. 

\subsection{Decentralized Training with Backup Workers}
\label{backup_concept}

An effective technique to mitigate the effect of heterogeneity is to use backup workers \cite{ICLR2016_backup_workers}. In centralized training, 
it can be easily implemented by allowing
the number of updates needed at parameter servers to 
be smaller than the number of workers.
Assume that there are $n$ workers and the number of updates needed in every iteration is $m$ ($m<n$). From 
each PS's perspective, the effective number of workers is $m$, since it needs $m$ updates in each iteration, and the remaining $(n-m)$ workers are ``backups''. In this way, we can tolerate at most $(n-m)$ slow workers in case of random slowdowns or even accidental node crashes without influencing the training speed. 

\begin{figure}
    \centering
    \includegraphics[width=\linewidth]{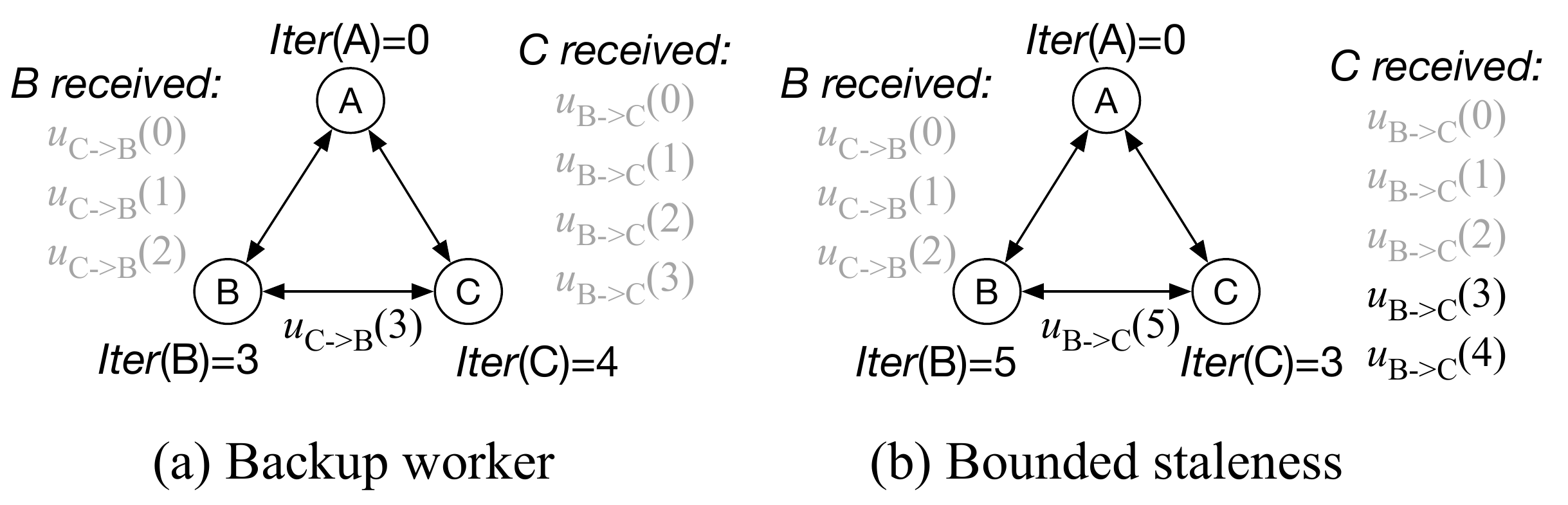}
    \caption{Backup workers and bounded staleness in distributed heterogeneous environments. Consumed updates are shown in grey. (a) Worker B and C can advance to any iteration, only relying on the updates of one another. (b) With the staleness bound set to 2, worker B is blocking due to the slow progress of A. Worker C can freely advance to the next iteration.}
    \label{fig:buwstaleEg}
\end{figure}

We naturally apply backup workers to decentralized training by setting the number of updates needed at each worker to be smaller than the number of its in-coming neighbors. As illustrated in Figure \ref{fig:buwstaleEg}(a), in a decentralized 3-worker setting, every worker has edges to and from two other workers, and only needs one update from its neighbors in every iteration. 
In the current state, worker A is stuck at iteration 0, while worker B and C are able to advance to iteration 3 and 4, respectively. Without backup workers, worker A would have dragged down the progress of B and C, because they would have both relied on A's updates to advance. With backup workers, B and C can advance to 
later iterations. 

However, the simple mechanism causes a fundamental problem: the iteration gap between two workers can be
{\em arbitrarily} large. 
It can be easily seen from Figure \ref{fig:buwstaleEg}, 
in which B and C can in fact be at {\em any}
iteration since they may only rely on the updates
of each other. 

Since the \textsc{notify-ack} mechanism 
implies that the iteration difference of adjacent 
workers is at most 2, 
the benefits of backup workers cannot be fully realized.  
For example, 
worker B and C can both be stuck at iteration 1 waiting for the $ACK(0)$ from A, which will not arrive without A's progress.
Thus, worker A can still drag down the progress
of worker B and C even if they do not need to wait
for worker A's update.
In essence, it is the {\em mechanisms} in \textsc{notify-ack}
that prevent the realization of backup workers. 

\subsection{Decentralized Training with Bounded Staleness}
\label{staleness_concept}

Bounded staleness~\cite{NIPS2013_SSP} is another
technique for centralized training to tolerate 
slow workers or slow communication between parameter
servers and workers. 
To realize it, an asynchronous parameter server is adopted but an upper-bound is enforced on the difference of the fastest worker's iteration and the slowest one's.
A worker is free to advance to a new iteration as long as the staleness bound is preserved. 

For decentralized training, it is difficult to enforce 
globally bounded staleness, which means that the iteration
difference of the fastest worker and slowest worker
in the whole system cannot exceed the bound. 
Clearly, it will defeat the decentralized nature
by introducing some kind of global progress monitor
to ensure such a property. 
Instead, we propose to apply bounded staleness in a local fashion: a worker can enter a new iteration as long as it has received updates from at most $s$ iterations ago from all its in-coming neighbors, where $s$ is the upper-bound of staleness.
Since updates are delivered locally, the enforcement of the staleness bound is straightforward. 
We believe that a local bound is a natural adoption of 
bounded staleness in the decentralized setting
that leads to efficient implementation.



Figure~\ref{fig:buwstaleEg}(b) shows an example of a staleness bound of 5 being used in the same 3-worker decentralized setting. Worker A, B and C are in the 0th, 5th and 3rd iteration, respectively. Worker A is temporarily slowed down due to a random factor, e.g., resource sharing. Nevertheless, with the use of staleness, worker B and C can continue to advance until the 5th iteration. The advantage of bounded staleness is that some progress can still be made even if certain workers are slowed down, --- the effect of the slowdown is mitigated. 

Let us consider \textsc{notify-ack} again.
Unfortunately, it directly imposes a strict bound on the iteration gap (i.e., 2), so
any staleness bound larger than 1 
is not possible. For example, in the case in Figure \ref{fig:buwstaleEg}(b), worker B and C would not have been able to enter iteration 1 
if \textsc{notify-ack} was used as the protocol, since they would still be waiting for $ACK(0)$ from worker A.
Similar to backup workers,
\textsc{notify-ack}'s mechanism affects the realization 
of local bounded staleness.

The {\bf essential takeaway} of the discussion so far is the 
following.
Although \textsc{notify-ack} points out and prevents 
the mixed-version problem, it does {\em not} realize the 
larger and potentially arbitrary iteration gap.
As we have shown, \textsc{notify-ack} is overly 
restrictive to force a very small gap between 
adjacent workers, which will 
{\em 1)} limit the potential of decentralized training; and
{\em 2)} prevent the implementation of backup workers and
bounded staleness.
To support larger iteration gaps while solving the mixed-version problem, we propose a queue-based coordination
scheme. 

\section{Queue-based synchronization}
\label{sec:queue}
This section first presents the update queue and the token queue
as the building blocks for decentralized training protocol.
Then, we present how they can be used to efficiently 
implement backup workers and bounded staleness.

\subsection{Update Queue}
\label{update}



To support mixed-version and large iteration gaps, 
we propose a queue-based coordination scheme where the received updates are stored in FIFO queues, called {\em update queues}. The update queue at worker $i$ is denoted by $UpdateQ(i)$. We further define the following queue operations:

\begin{itemize}[leftmargin=*]
    \item $q.enqueue(update,iter=None,w\_id=None)$ pushes $update$ into the queue, where $iter$ and $w\_id$ are tags, denoted by $(iter,w\_id)$, and $q$ is the name of the queue. The input $iter$ indicates the index of the iteration where the $update$ was generated, and $w\_id$ indicates the index of the sender worker.
    
    \item $q.dequeue(m,iter=None,w\_id=None)$ takes the first $m$ entries tagged with $(iter,w\_id)$ out of the queue and returns a list containing these entries. This function blocks if there are not enough elements tagged with $(iter,w\_id)$ in the queue. If one of the tags is not specified, then the first $m$ entries matching the other tag will be returned. If neither is specified, the first $m$ entries are returned regardless of their tags. If needed, tags of the returned entries can be returned as well.
    
    \item $q.size(iter=None,w\_id=None)$ returns the number of entries tagged with $(iter,w\_id)$ in the queue. If one of the tags is not specified, the number of entries matching the other tag is returned. If neither is specified, the total number of entries in the queue is returned.
\end{itemize}

\begin{figure}
    \centering
    \includegraphics[width=\linewidth]{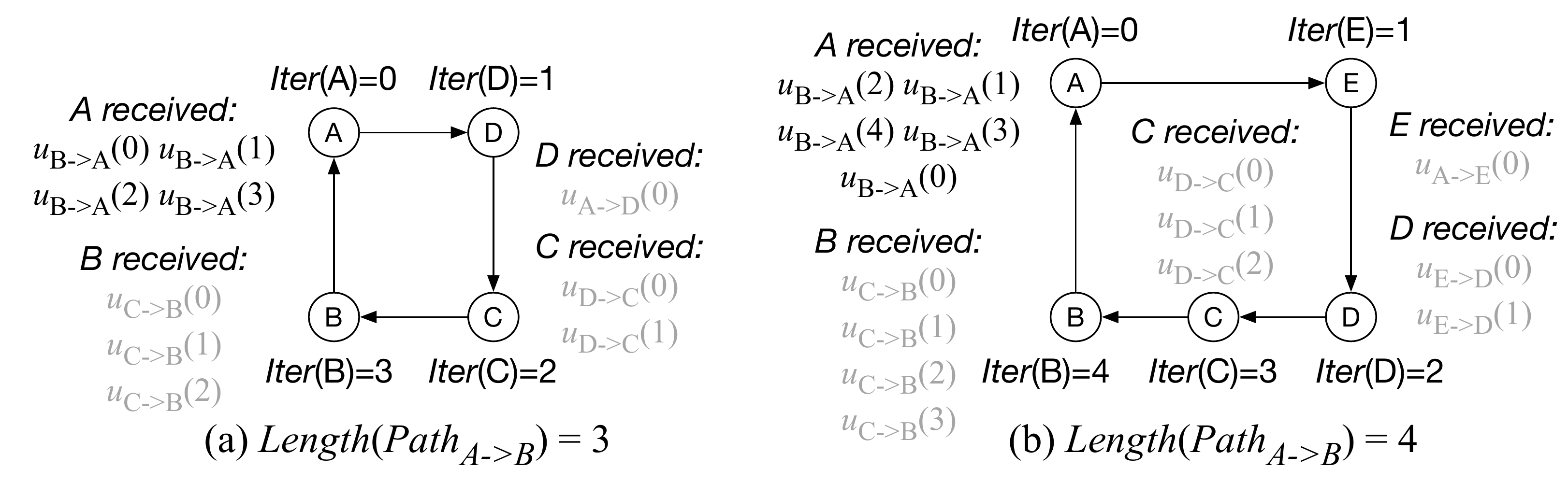}
    \caption{Iteration gap in standard decentralized training. The size of the update queue is 4. Consumed updates are shown in grey. (1) A maximum iteration gap of 3 between B and A is illustrated. $UpdateQ(A)$ is able to accommodate the 4 updates. (2) A maximum iteration gap of 4 between B and A is observed. $UpdateQ(A)$ will not be able to accommodate the updates. Using $TokenQ(A \rightarrow B)$ can prevent this situation.}
    \label{fig:standardDec}
\end{figure}

Based on the update queue, the standard decentralized training algorithm is shown in Figure \ref{alg1}. 
To send an update from $i$ to $j$, 
worker $i$ directly enqueues the parameters to the update queue
of worker $j$.
To receive an update, a worker can locally dequeue 
updates sent from various workers and iterations.
However, one question remains, --- how large should the queue be to accommodate all the updates? 
Based on \textit{Theorem 1}, for any worker $i$ and its in-coming neighbor $j$, worker $j$ can be $length(Path_{i \rightarrow j})$ iterations ahead of $i$.
It means that $UpdateQ(i)$ must be able to store updates of $length(Path_{j \rightarrow i}) + 1$ different iterations from $j$. When the number of workers is large, the shortest path from $j$ to $i$ can also be large, and so must be the capacity of the queue, which will then put considerable pressure on the system memory. An example is shown in Figure \ref{fig:standardDec}.

\begin{figure}
\centering
\small
 \begin{algorithmic}[1]
 \REQUIRE Initial model parameters $p_0$
 \REQUIRE $UpdateQ(i)$ for all $i \in V$
 \REQUIRE Maximum number of iterations $max\_iter$
 \FOR {$i \in V$}
 \STATE //Initialize local model parameters
 \STATE $x_{0,i} = p_0$
 \FOR {$k = 0$ \TO $max\_iter$}
 \STATE // 1. $Send$ my parameters to my out-going neighbors
 \FOR {$j \in N_{out}(i)$}
 \STATE $UpdateQ(j).enqueue(x_{k,i},iter=k,w\_id=i)$
 \ENDFOR
 \STATE // 2. $Compute$ gradients based on $x_{k,i}$
 \STATE Randomly sample a batch of data $d_{k,i}$
 \STATE $grads = Compute(x_{k,i},d_{k,i})$
 \STATE // 3. $Recv$ parameters from my in-coming neighbors
 \STATE $x_{recv} = UpdateQ(i).dequeue(|N_{in}(i)|,iter=k)$
 \STATE // 4. $Reduce$
 \STATE $temp = \sum_{j = 0}^{|N_{in}(i)|-1} x_{recv}(j) / |N_{in}(i)|$
 \STATE // 5. $Apply$
 \STATE $x_{k+1,i} = temp+grads$
 \ENDFOR
 \ENDFOR
 \end{algorithmic}
\caption{Decentralized Training with Update Queue}
\label{alg1}
\end{figure}

\subsection{Token Queue: Controlled Iteration Gaps}
\label{token}

To tackle this problem, we propose token queues as a mechanism to {\em control the iteration gap between adjacent workers}. Note that by the nature of standard decentralized training, every worker can be at most one iteration {\em ahead} of its in-coming neighbors; therefore, we only need to control a worker's speed as compared to its potentially {\em slower} out-going neighbors. 

In our design, each worker maintains a token queue for every in-coming neighbor. Whenever a worker attempts to enter a new iteration, it must acquire a token from every one of its out-going neighbors. The number of tokens in the queue determines how many more iterations an in-coming neighbor can advance, considering the local worker's progress. Assuming that we want the iteration gap between adjacent workers not to exceed a predefined positive integer constant $max\_ig$,
we propose the following procedure to ensure this gap. 
We denote the token queue at worker $i$ storing tokens for worker $j$ by $TokenQ(i \rightarrow j)$, where $i \in N_{out}(j)$.

\begin{itemize}[leftmargin=*]
    \item {\bf Initialization}
    At the start of the first iteration, each worker puts $max\_ig$ tokens in each token queue it maintains.
    
    \item {\bf Remove token}
    When a worker $i$ attempts to enter a new iteration, it must remove a token from {\em every} one of its out-going neighbors, 
    --- for each $j \in N_{out}(i)$, remove a token from 
    $TokenQ(j \rightarrow i)$.
    
    \item {\bf Insert token}
    When a worker $i$ enters a new iteration, it will insert a token in {\em every} local token queue, --- for each $j \in N_{in}(i)$, insert a token to $TokenQ(i \rightarrow j)$. This allows all its in-coming neighbors to advance further.
\end{itemize}

\textbf{Theorem 2}. For standard decentralized training, with token queues, the upper-bound of $Iter(i)-Iter(j)$ is given by:

$min(length(Path_{j \rightarrow i}),max\_ig \times length(Path_{i \rightarrow j}))$.

\begin{figure}
    \centering
    \includegraphics[width=\linewidth]{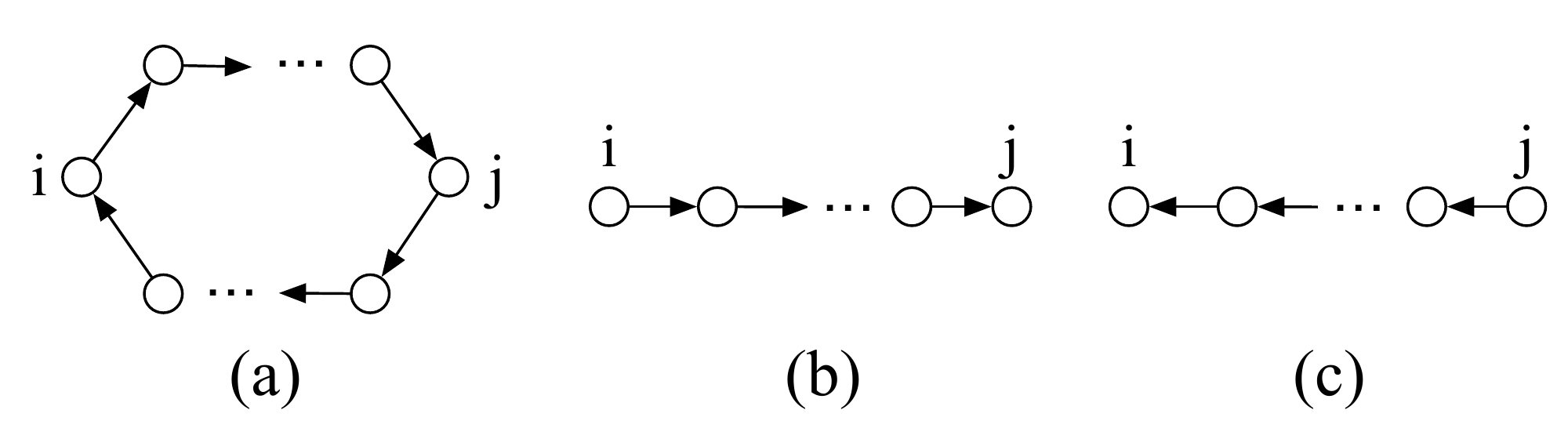}
    \caption{Topological relations between two workers.}
    \label{token_proof}
\end{figure}

\textit{Proof}.  First, we consider a pair of adjacent workers $(i,j)$. For any worker $i$ and its out-going neighbor $j$, we already know from \textit{Theorem 1} that $Iter(j)-Iter(i) \leq 1$. To derive the upper-bound of $Iter(i)-Iter(j)$, we will prove in a deductive manner that $TokenQ(j \rightarrow i).size() = Iter(j)-Iter(i)+max\_ig$ holds for all iterations, where the $size()$ function returns the number of tokens in the token queue. At the start of training, $Iter(i) = Iter(j) = 0$, and the initialization of $TokenQ(j \rightarrow i)$ results in $TokenQ(j \rightarrow i).size() = max\_ig$. Therefore, the above equation holds. The variables in the equation only change when either worker $i$ or $j$ advances to a new iteration. If worker $i$ advances, it must remove one token from $TokenQ(j \rightarrow i)$; at the same time, $Iter(i)$  is increased by 1. Therefore, both sides of the equation are decreased by 1 and the equality still holds. Similarly, if worker $j$ enters a new iteration, it must insert one token into $TokenQ(j \rightarrow i)$; therefore, both sides of the equation are increased by 1 and the equality still holds. Since the number of tokens in $TokenQ(j \rightarrow i)$ is non-negative, we have $Iter(j)-Iter(i)+max\_ig = TokenQ(j \rightarrow i).size() \geq 0$, and thus $Iter(i)-Iter(j) \leq max\_ig$.

Next we consider an arbitrary pair $(i,j)$. Since the graph is connected, there exists a path from $i$ to $j$ and vice versa, as shown in Figure~\ref{token_proof} (a).
The two paths can be viewed as two basic scenarios. 
In Figure~\ref{token_proof} (b), 
due to the earlier proof of the 
adjacent workers case, 
$Iter(i)-Iter(j) \leq max\_ig \times length(Path_{i \rightarrow j})$.
In Figure~\ref{token_proof} (c), due to {\em Theorem 1},
$Iter(i)-Iter(j) \leq length(Path_{j \rightarrow i})$.
The general case is the combination of the two
scenarios, thus we have 
$Iter(i)-Iter(j) \leq min(length(Path_{j \rightarrow i}),max\_ig \times length(Path_{i \rightarrow j}))$.
$\blacksquare$

The intuition of the iteration gap being
bounded by the smaller one is that, otherwise, the 
larger gap becomes {\em infeasible}, due to 
either not having enough tokens 
(if the gap in Figure~\ref{token_proof} (b) is larger), or
not actually having long enough path
(if the gap in Figure~\ref{token_proof} (c) is larger).
Overall, the proposed token queue provides a flexible parametrized method to bound the iteration gap.  
The upper-bound of the capacity of any token queue is
$max\_ig \cdot (length(Path_{i \rightarrow j})+1)$.
It directly follows from applying the upper-bound of $Iter(i)-Iter(j)$ proved in \textit{Theorem 2} to $TokenQ(i \rightarrow j).size()=Iter(i)-Iter(j)+max\_ig$.


Now we apply token queues to the example in Figure \ref{fig:standardDec}(b). We can set $max\_ig$ to 3. The token queue at worker A contains 3 tokens at the beginning of the 0th iteration. Whenever worker B enters a new iteration, it must obtain a token from A. Since A has not progressed, B can get at most 3 tokens from A, which enables B to reach the 3rd iteration but no more. Therefore, A only has to deal with at most 4 updates at a time, and the situation in the figure is prevented.
 
The decentralized training algorithm 
using token queues is shown in Figure \ref{alg2}. With bounded iteration gaps, the required size of $UpdateQ(i)$ is upper-bounded by $(1+max\_ig)|N_{in}(i)|$, regardless of the graph size or topology. 
Although the use of token queues may 
only provide a marginal improvement to the algorithm
based merely on the update queue,
we will later see in Section~\ref{backup} 
that bounding the iteration gap is absolutely necessary when backup workers are employed to mitigate the effect of heterogeneity.

\begin{figure}
\centering
\small
 \begin{algorithmic}[1]
 \REQUIRE All the requirements in Figure \ref{alg1}
 \REQUIRE $TokenQ(i \rightarrow j)$ for all $i \in V$ and all $j \in N_{in}(i)$
 \REQUIRE Maximum iteration gap $max\_ig$
 \FOR {$i \in V$}
 \STATE $x_{0,i} = p_0$
 \FOR {$j \in N_{in}(i)$}
 \STATE // Put $(max\_ig-1)$ initial tokens
 \STATE $TokenQ(i \rightarrow j).enqueue([0] *(max\_ig-1))$
 \ENDFOR
 \FOR {$k = 0$ \TO $max\_iter$}
 \FOR {$j \in N_{in}(i)$}
 \STATE // Insert tokens
 \STATE $TokenQ(i \rightarrow j).enqueue([k])$
 \STATE $Send(x_{k,i},k,i)$ \algorithmiccomment{1. $Send$}
 \STATE $grads = Compute(x_{k,i})$ \algorithmiccomment{2. $Compute$}
 \STATE $x_{recv} = Recv(k,i)$ \algorithmiccomment{3. $Recv$}
 \STATE $temp = Reduce(x_{recv})$ \algorithmiccomment{4. $Reduce$}
 \STATE $x_{k+1,i} = temp+grads$ \algorithmiccomment{5. $Apply$}
 \FOR {$j \in N_{out}(i)$}
 \STATE // Get a new token
 \STATE $TokenQ(j \rightarrow i).dequeue(1)$
 \ENDFOR
 \ENDFOR
 \ENDFOR
 \ENDFOR
 \end{algorithmic}
\footnotesize
\emph{Notes:} Some pseudocodes have been wrapped up in functions. Line 11 has replaced the original lines 6-8 from Figure \ref{alg1}; line 12 has replaced the original lines 10-11; line 13 has replaced the original line 13; line 14 has replaced the original line 15.
\caption{Decentralized Training with Token Queue}
\label{alg2}
\end{figure}

\subsection{Supporting Backup Workers}\label{backup}

Applying backup workers to decentralized training is relatively intuitive. As stated in Section~\ref{backup_concept}, instead of requiring an update from every in-coming neighbor, a worker only needs to get updates from a few neighbors in order to advance to the next iteration, i.e., the number of updates needed is smaller than the number of its in-coming neighbors. In the algorithm shown in Figure \ref{alg4}, when collecting updates from the local update queue, a worker first makes sure it has enough number of updates by specifying the number in the input of the $dequeue$ function. Then it checks the queue for any additional update that is available, as the number of updates received may exceed the required number. The above process can also be replaced with a while loop that continues to take available updates out of the queue until the amount is adequate.

One problem with using a smaller number of updates is that the unused updates that arrive later can accumulate in the update queue. Iteration after iteration, they will take up more and more amount of space, which will inevitably lead to overflow. We propose a solution that consists of two parts: 
{\em a)} clear the stale updates periodically; and 
{\em b)} with little communication overhead, prevent unnecessary updates from being sent by checking the receiver's iteration before the $Send$. We will explain in more detail in Section~\ref{sec:impl}.

Another distinct feature of the backup workers setting is that the iteration gap is unbounded. As we have illustrated in Figure \ref{fig:buwstaleEg} before, worker B and C can progress to an arbitrarily large iteration depending only on the updates of one another, while their common neighbor A stays in iteration 0. Therefore, bounding the iteration gap is a {\em must} for the correct execution of decentralized training --- in this case, token queues are an indispensable part of the design.

\begin{figure}
\centering
\small
 \begin{algorithmic}[1]
 \REQUIRE All the requirements in Figure \ref{alg2}
 \REQUIRE Number of backup workers $N\_buw(i)$ for all $i \in V$; $N\_buw(i) < |N_{in}(i)|$
 \STATE function $Recv(k,i)\{$
  \STATE\hspace{\algorithmicindent} // Get the needed updates
  \STATE\hspace{\algorithmicindent} $x_{rcv1} = UpdateQ(i).dequeue(|N_{in}(i)|-N\_buw(i),iter=k)$
  \STATE\hspace{\algorithmicindent} // Get additional updates remaining in the queue
  \STATE\hspace{\algorithmicindent} $x_{rcv2} = UpdateQ(i).dequeue(UpdateQ(i).size(k),iter=k)$
  \STATE\hspace{\algorithmicindent} // Combine updates and return
  \STATE\hspace{\algorithmicindent} \textbf{Return} $concatenate(x_{rcv1},x_{rcv2})$
 \STATE $\}$
 \STATE function $Reduce(x_{recv})\{$
  \STATE\hspace{\algorithmicindent} // Compute the number of entries in $x_{recv}$
  \STATE\hspace{\algorithmicindent} $N\_updates = size(x_{recv})$
  \STATE\hspace{\algorithmicindent} // Reduce and return
  \STATE\hspace{\algorithmicindent} \textbf{Return} $\sum_{j = 0}^{N\_updates-1} x_{recv}(j) / N\_updates$
 \STATE $\}$
 \STATE $Train()$
 \end{algorithmic}
\footnotesize
\emph{Notes:} Lines 1-22 in Figure \ref{alg2} are wrapped up in the function $Train()$.
\caption{Decentralized Training with Backup Workers}
\label{alg4}
\end{figure}

\subsection{Supporting Bounded Staleness}
\label{staleness}



As discussed in Section~\ref{staleness_concept}, in the bounded staleness setting, a worker can enter a new iteration as long as it has received updates from at most $s$ iterations ago from all its in-coming neighbors, where $s$ is the upper-bound on staleness. However, a specific way to incorporate staleness in decentralized training is yet to be discovered. In particular, it remains a problem how to handle stale updates. 

We make an observation that model parameters sent in a later iteration contain the information carried by earlier updates, since later updates are built upon earlier ones and gradients are accumulated in the parameters sent in the updates. Therefore, we propose to use the most recent available updates whenever possible and discard the rest. 
Specifically, when a worker collects updates from its local update queue, it will compare the tags and select the newest update from each of its in-coming neighbors. If the update is within the staleness bound, it is deemed satisfactory; otherwise, it is dropped. If no update from an in-coming neighbor, either received in the current iteration or in previous ones, is within the current staleness bound, the worker will block until it gets a newer update from the corresponding neighbor. When the worker has received a satisfactory update from every one of its in-coming neighbors, it will perform a $Reduce$ on the newly received updates. Note that the updates to reduce may come from different iterations, thus a simple average may not be the best way to aggregate them. We have compared simple averaging to an iteration based weighted average, and found the latter performs slightly better. For worker $j$ in iteration $k$, the update formula we have settled on is as follows:
\begin{equation}
\vspace{-2mm}
    \frac{\sum_{i \in N_{in}^{(k)}(j)} [Iter(u_i)-(k-s)+1]u_i}{\sum_{i \in N_{in}^{(k)}(j)} [Iter(u_i)-(k-s)+1]}
\label{eq:updateST}
\end{equation}
where $Iter(u_i)$ is the iteration in which $u_i$ was generated, and $N_{in}^{(k)}(j) =$\{$i \in N_{in}(j)$: $u_i$ received in iteration $k$ is satisfactory\}. The weight of an update is linearly associated with its iteration, which is at least $k-s$ to be considered satisfactory. The above formula may very well be non-optimal, and we leave further 
optimization as future work.

As for the iteration gap, with a staleness bound of $s$, we have $Iter(i) - Iter(j) \leq s+1$ for $j \in N_{in}(i)$. This is because for worker $i$ to enter $Iter(i)+1$, it needs an update from worker $j$ at least as recent as $u_{j \rightarrow i}(Iter(i)-s)$. Therefore, the upper-bound on the iteration gap is given by
$Iter(i) - Iter(j) \leq (s+1) \cdot length(Path_{j \rightarrow i})$.
We see that the iteration gap has been largely increased compared to the standard decentralized setting. Therefore, we have also employed token queues as a way to bound the gap. The algorithm is shown in Figure \ref{alg3}.

\begin{figure}
\centering
\small
 \begin{algorithmic}[1]
 \REQUIRE All the requirements in Figure \ref{alg2}
 \REQUIRE Staleness bound $max\_staleness$
 \REQUIRE The iteration of the most recent $u_{i \rightarrow j}$ received, denoted by $iter\_rcv_{i \rightarrow j}$, initialized to -1, for $i \in V$ and $j \in N_{out}(i)$
 \STATE function $Recv(k,i)\{$
 \STATE\hspace{\algorithmicindent} $min\_iter = k-max\_staleness$
 \STATE\hspace{\algorithmicindent} $x_{recv} = [], iter_{recv} = []$
 \STATE\hspace{\algorithmicindent} \textbf{for} $j \in N_{in}(i)$ \textbf{do}
 \STATE\hspace{\algorithmicindent}\hspace{\algorithmicindent} \textbf{do}
 \STATE\hspace{\algorithmicindent}\hspace{\algorithmicindent}\hspace{\algorithmicindent} $q\_sze = max(UpdateQ(i).size(w\_id=j),1)$
 \STATE\hspace{\algorithmicindent}\hspace{\algorithmicindent}\hspace{\algorithmicindent} $(l\_x,l\_iter) = UpdateQ(i).dequeue(q\_sze,w\_id=j)$
 \STATE\hspace{\algorithmicindent}\hspace{\algorithmicindent}\hspace{\algorithmicindent} $iter\_rcv_{j \rightarrow i}=max(max(l\_iter),iter\_rcv_{j \rightarrow i})$
 \STATE\hspace{\algorithmicindent}\hspace{\algorithmicindent} \textbf{while} $iter\_rcv_{j \rightarrow i}<min\_iter$
 \STATE\hspace{\algorithmicindent}\hspace{\algorithmicindent} \textbf{if} $max(l\_iter) \geq min\_iter$ \textbf{then}
 \STATE\hspace{\algorithmicindent}\hspace{\algorithmicindent}\hspace{\algorithmicindent} $x_{recv} = concatenate(x_{recv},l\_x(argmax(l\_iter)))$
 \STATE\hspace{\algorithmicindent}\hspace{\algorithmicindent}\hspace{\algorithmicindent} $iter_{recv} = concatenate(iter_{recv},[max(l\_iter)])$
 \STATE\hspace{\algorithmicindent}\hspace{\algorithmicindent} \textbf{endif}
 \STATE\hspace{\algorithmicindent} \textbf{end for}
 \STATE\hspace{\algorithmicindent} // Return a tuple of the parameters, their iterations and $k$
 \STATE\hspace{\algorithmicindent} \textbf{Return} $tuple(x_{recv},iter_{recv},k)$
 \STATE $\}$
 
 \STATE function $Reduce(tuple_{recv})\{$
 \STATE\hspace{\algorithmicindent} // Deconstruct the input tuple
 \STATE\hspace{\algorithmicindent} $(x_{recv}, iter_{recv}, k) = tuple_{recv}$
 \STATE\hspace{\algorithmicindent} // Compute the number of entries in $x_{recv}$
 \STATE\hspace{\algorithmicindent} $N\_updates = size(x_{recv})$
 \STATE\hspace{\algorithmicindent} // Reduce the updates
 \STATE\hspace{\algorithmicindent} $sum\_weight = \sum_{j = 0}^{N\_updates-1} [iter_{recv}(j)-(k-s)+1]$
 \STATE\hspace{\algorithmicindent} $temp = \frac{\sum_{j = 0}^{N\_updates-1} [iter_{recv}(j)-(k-s)+1]x_{recv}(j)}{sum\_weight}$
 \STATE\hspace{\algorithmicindent} \textbf{Return} $temp$
 \STATE $\}$
 \STATE $Train()$
 \end{algorithmic}
\footnotesize
\emph{Notes:} Lines 1-22 in Figure \ref{alg2} are wrapped up in the function $Train()$.
\caption{Decentralized Training with Bounded Staleness}
\label{alg3}
\end{figure}

\begin{table*}[t]
\small
\begin{center}
\begin{tabular}[t]{|l|l|l|l|}
\hline
Setting & For $j \in N_{in}(i)$ & For $i \in N_{in}(j)$ & For arbitrary $(i,j)$  \\
 \hline
Standard decentralized & $1$ & $length(Path_{j \rightarrow i})$ & $length(Path_{j \rightarrow i})$ \\
\hline
Bounded staleness  & $s+1$ & $(s+1) \times length(Path_{j \rightarrow i})$ & $(s+1) \times length(Path_{j \rightarrow i})$ \\
\hline
Backup worker & $\infty$ & $\infty$ & $\infty$ \\
\hline
Hybrid & $\infty$ & $\infty$ & $\infty$ \\
\hline
Using \textsc{notify-ack} & $1$ & $2$ & $min(length(Path_{j \rightarrow i}),2 \times length(Path_{i \rightarrow j}))$ \\
\hline
Using token queues & $b_0$ (varied) & $max\_ig$ & $min(b_0 \times length(Path_{j \rightarrow i}),max\_ig \times length(Path_{i \rightarrow j}))$ \\
\hline
\end{tabular}
\caption{Theoretical upper-bound on the iteration gap $Iter(i)-Iter(j)$ for various settings. $b_0$ is varied according to the original setting to which token queues are applied, e.g. $b_0=1$ for standard decentralized setting and $b_0=s+1$ for bounded staleness. For backup worker and the hybrid setting, the original bound is $\infty$, therefore $b_0$ can only be derived from the last column, which gives $b_0 = max\_ig \cdot length(Path_{i \rightarrow j})$.  However, no matter the setting to which token queues are applied, the maximum number of tokens in a token queue is always $TokenQ(i \rightarrow j).size() \leq max\_ig \cdot (length(Path_{i \rightarrow j})+1)$.
\label{tbl:ig_ubnd}}
\end{center}
\end{table*}

\section{Deterministic Slowdown}
\label{sec:opt4het}
To mitigate the effect of heterogeneity, backup workers and bounded staleness loosen up the synchronization scheme by allowing a worker to advance faster than a few or even all of its incoming neighbors. However, the gap between a worker's iteration and its neighbor's is still bounded, either by the technique itself (in the bounded staleness case) or by the use of token queues (in the backup worker case). Consequently, if a worker suffers from severe deterministic slowdown, it will eventually drag down its neighbors and then the entire graph of nodes. Therefore, a dilemma exists: if we do not bound the iteration gap, then better performance can be achieved against deterministic slowdown, but such a system is unrealistic, since a worker must be prepared to receive updates from infinitely many iterations; on the other hand, if we bound the iteration gap with mechanisms like token queues, then whenever a node is slowed down
in a deterministic manner, other nodes can only obtain limited progress before having to wait for the slow worker, and the maximum progress they can make is strictly determined by the bound shown in the last row of Table \ref{tbl:ig_ubnd}.

There are two potential solutions. 
One is 
developing new {\em algorithms}
to eliminate the bound on the iteration gap and support infinitely large iteration gaps.
In fact, AD-PSGD~\cite{ArXiv_ADPSGD} is one example, where every worker updates its own parameters by averaging them with a randomly selected neighbor at the end of each iteration, regardless of the neighbor's iteration. However, this algorithm easily creates deadlock, and to prevent it, existing solutions require the communication graph $G$ to be bipartite \cite{ArXiv_ADPSGD}, which greatly constrains users' choice of communication topology. 
The other is developing new {\em system}
mechanisms to identify the slow worker and seek a way to let it progress, so that the training process can resume.

We consider a system approach. 
We let the slow worker identify itself by checking the number of tokens in its out-going neighbors' token queues. This can be done conveniently when it acquires the needed tokens from its out-going neighbors at the end of each iteration, in order to enter a new iteration. For worker $i$ and its out-going neighbor $j$, the number of tokens in $TokenQ(j \rightarrow i)$ is exactly $Iter(j)-Iter(i)+max\_ig$. If $TokenQ(j \rightarrow i).size()$ is large for all $j \in N_{out}(i)$, we can imagine that worker $i$ is very likely a straggler. 

\begin{figure}[b]
    \centering
    \vspace{-6mm}
    \includegraphics[width=\linewidth]{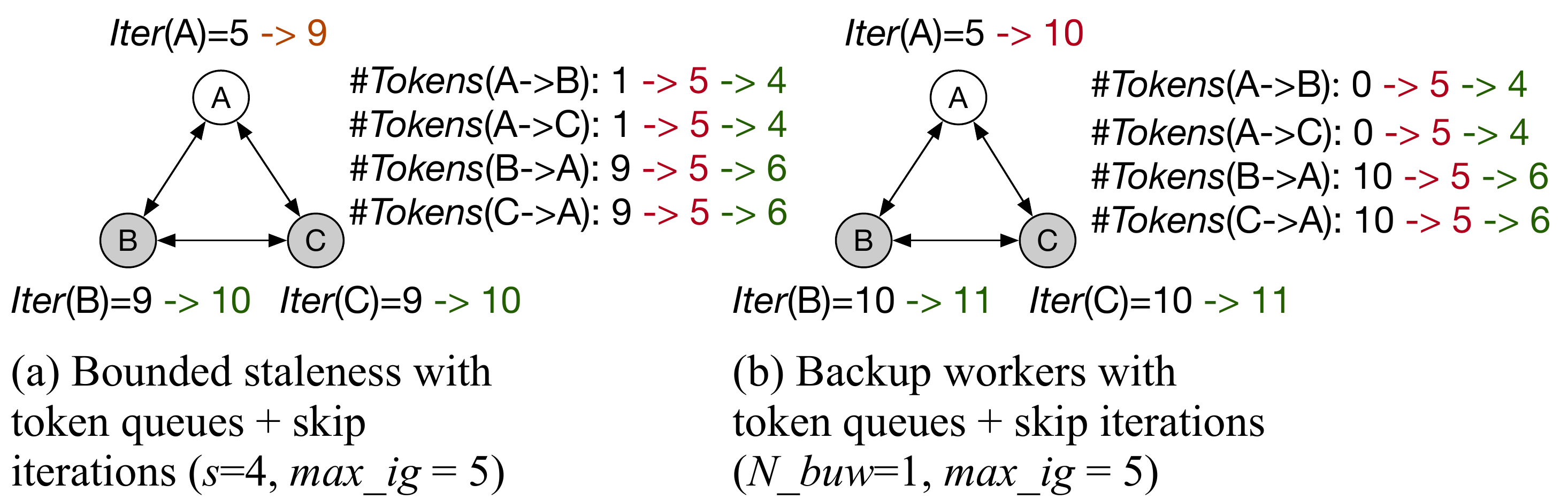}
    \caption{Skipping Iteration Example. \#Tokens denotes the number of tokens in the corresponding token queue. }
    \label{fig:JI}
\end{figure}

To allow the straggler to make progress, we propose {\em skipping iterations}, i.e., a slow worker can jump a few iterations, allowing other workers to advance. According to the token queues scheme proposed in Section~\ref{token}, the maximum number of iterations worker $i$ can jump is determined by $max\_jump = \min_{j \in N_{out}(i)} TokenQ(j \rightarrow i).size()$, since every new iteration it enters, it needs a new token from every one of its out-going neighbors. Although skipping $max\_jump$ iterations is allowed under our proposed token queues scheme, we argue that it would be absurd for the slow worker $i$ to surpass its out-going neighbors after the jump; therefore, a more intuitive upper-bound is given by $max\_jump-max\_ig$. 

Assume that worker $i$ will jump to iteration $k$. In our design, before the jump, worker $i$ will renew its parameters by executing $Recv(k-1)$ and a following $Reduce$, averaging its current parameters with updates sent by its in-coming neighbors in the $(k-1)$-th iteration, i.e. $\{u_{j' \rightarrow i}(k-1): j' \in N_{in}(i) \}$. This is to ensure that after the jump, worker $i$'s parameters will not appear too stale, so that future updates sent by worker $i$ will not harm the training process. In this way, the slow worker will not always remain a straggler. Note that to ensure the correctness of the token queues scheme, when a jump from iteration $k_0$ to iteration $k$ is performed, the worker will need to obtain $(k-k_0)$ tokens from every one of its out-going neighbors, and also put $(k-k_0)$ tokens into every local token queue intended for its in-coming neighbors. 

Figure \ref{fig:JI} gives two examples of executing the jump, one for the bounded staleness case and the other for backup workers. Changes in red indicate the jump, while changes in green indicate the new progress enabled by the jump. (a) Worker B and C are blocked from advancing because of the staleness bound of 4. Without A's skipping iterations, the speed of B and C will be no more than 4 iterations faster than A. However, with skipping iterations, the slow worker A can quickly jump to iteration 9, so that training can smoothly resume for another 4 iterations, before A advances again. (b) Worker B and C are blocked because of the bounded iteration gap ensured by the token queues --- they cannot get a new token from A, since the token queues are empty. Without A's skipping iterations, they will be no more than 5 iterations faster than A. But with A jumping to the 10th iteration, they can train non-stop for another 5 iterations, before A makes new progress. 

It may seem like a problem that for worker $i$ to execute $Recv(k-1)$, it must wait for its in-coming neighbors to reach the $(k-1)$-th iteration. However, we argue that it is very likely that worker $i$ has also fallen behind its in-coming neighbors due to the deterministic slowdown; even if one of its in-coming neighbors is also slow, the mechanism of either bounded staleness or backup workers will ensure that worker $i$ can easily proceed to the $k-th$ iteration. Moreover, although skipping iterations means missing a few iterations' updates from worker $i$, this will not be a problem because even if the updates were sent, they would be stale and thus dropped by worker $i$'s out-going neighbors (in the backup workers case) / receive a very small weight (in the bounded staleness case).

To enable flexible settings of our proposed mechanism, our system allows users to specify the maximum number of iterations that a worker can skip in one jump, as well as the condition to trigger the jump, e.g., a worker may only skip iterations if it is more than a user-specified number of iterations behind its out-going neighbors.

\section{Implementation}
\label{sec:impl}
We implemented our system in \textsc{TensorFlow}. 
Specifically, the queues in our design 
are based on the comprehensive and flexible FIFO queue in \textsc{TensorFlow}. 
The initialization specifies the data types, shapes and the capacity of the queue entries. 
The FIFO queue supports common 
functions including $enqueue$, $dequeue$,
$dequeue\_many$ and $size$.


\subsection{Collecting Updates Matching a Tag}
\label{sec:implTag}

To implement the queue operations defined in Section~\ref{update}, we only need to enhance
each FIFO queue entry with a tag, which is 
used to match an update of a particular iteration and/or from a particular neighbor.

A simple implementation of the tags is to use one FIFO queue as the Update Queue at each worker and include the tags as part of the queue entry. Whenever a worker collects updates from the queue, it takes all entries out of the queue and keeps the ones with matching tags. This process goes on in a while loop until the worker has obtained all the required entries. The issue with this approach is that dealing with the unmatched 
entries can be cumbersome. They cannot be discarded since
they can be from later iterations and will be used in the future. This may happen because 
we do not assume network preserves the 
message order. 
We cannot simply put them back into the queue, since they will be dequeued again and again as the while loop continues. It is possible to store them locally after performing a partial $Reduce$ of the available updates according to the tag, but that will complicate the bookkeeping and
consume a considerable amount of local memory, --- about $max\_ig$ times model size.

We propose a solution that prevents dequeuing updates of newer iterations with nearly zero memory overhead. Instead of using a single queue, we define multiple queues, each of which corresponds to an iteration. Queues are reused across iterations in a way similar to rotating registers. To select the correct queue to $enqueue$ or $dequeue$, a worker determines the queue index by computing the modulo $mod(iter,\#queues)$, where $\#queues$ is the total number of queues. $\#queues$ is set to $max\_ig+1$, because a worker can receive updates of at most $(max\_ig+1)$ different newer or current iterations based on {\em Theorem 1}. 
In standard case (Section~\ref{update}),
a worker can only receive newer or current updates, and it can always $dequeue$ the correct updates;
in the backup worker case (Section~\ref{backup}),
a worker can receive older updates as well, but the older ones will be discarded.
Our solution essentially divides the original large single queue into multiple small ones, and the total space consumed basically remains the same.

As for distinguishing the sender via the $w\_id$ tag, it can also be achieved by defining multiple queues. But it is not necessary in our system, since we only use the $w\_id$ tag when employing bounded staleness, which only requires processing
one-pass of all entries: among the entries with the same $w\_id$ tag, the most recent one is retained and the rest are discarded.

\subsection{Handling Late Updates}

As mentioned in Section~\ref{backup}, when using backup workers, updates that are not used in the $Reduce$ can accumulate in the queue. In our design, the effect of stale updates is mitigated in the following two ways:
{\em a)} Stale updates are found and discarded in the $dequeue$/$dequeue\_many$ operation in later iterations. This is already incorporated in our system as described in Section~\ref{sec:implTag}. 
{\em b)} Inquire the receiver's iteration before sending the update. If the receiver's iteration is larger than the local worker's iteration, do not send the update. This method creates a small communication overhead, but can save much more when the update is stale. More importantly, it can effectively reduce the number of stale updates --- now the only source of stale updates are those that are on-the-fly when the receiver performs the $dequeue$/$dequeue\_many$. 

We have also considered a more customized structure provided by \textsc{TensorFlow} called the conditional accumulator, which only accepts updates sent with a correct $local\_step$, another notion for iteration. If the $local\_step$ is incorrect, then the update will be dropped. It seemed to be a perfect solution to the problem, but we have observed in experiments that this property cannot be always ensured. The update that is up-to-date when it is sent can end up stale when it is received, and the conditional accumulator will incorrectly accept the update. This is exactly the same problem we have encountered with FIFO queues 
with the on-the-fly stale updates.

\section{Experiments}
\label{sec:exprmnts}
\subsection{Dataset and Models}

We evaluate \projectname on two machine learning tasks, namely image classification and web spam detection. For image classification we train the VGG11\cite{VGG11} network on CIFAR-10\cite{Cifar10}; for web spam detection, we train SVM over the webspam dataset\cite{pascalchallenge}.

\subsection{Experiments Setup}
We use a CPU cluster with 1000Mbit/s ethernet connection to run 16 workers on 4 machines: each machine has 4 workers.
We use the following hyper-parameter setup as prescribed in http://leon.bottou.org/projects/sgd and AD-PSGD \cite{ArXiv_ADPSGD} with some modifications:
batch size: 128; learning rate: 0.1 for VGG and 10 for SVM; no learning rate decay policy is used; 
momentum: 0.9; weight decay: $10^{-4}$ for VGG and $10^{-7}$ for SVM.
We use log loss for SVM instead of hinge loss.


\subsection{Results and Analysis}

\subsubsection{Heterogeneity with Random Slowdown}


\begin{figure}
    \centering
    \includegraphics[width=0.8\linewidth]{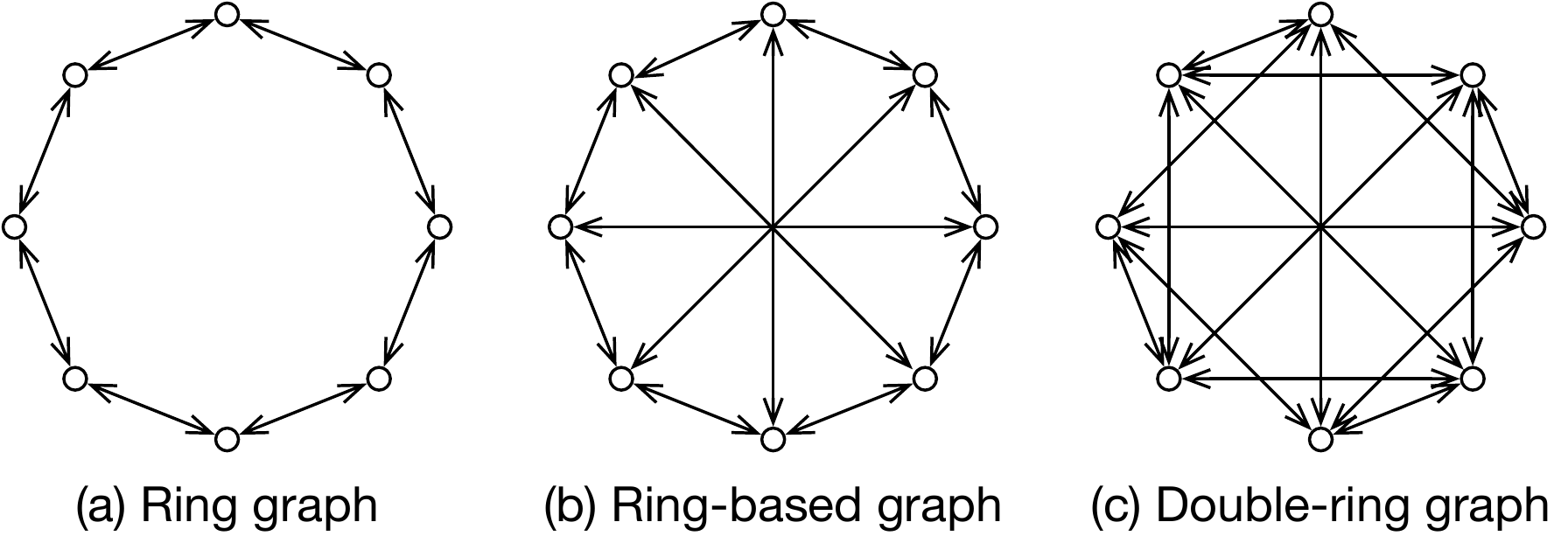}
    \caption{Graphs used in experiments (self-loops are omitted) with increasing node degrees. (1) Ring graph \cite{ArXiv_ASAP,NIPS2017_dPSGD,ArXiv_ADPSGD}. Nodes are connected in a circle via bidirectional edges. (2) A ring-based graph \cite{NIPS2017_dPSGD}. On top of the ring graph, every node is also connected to the most distant node. (3) Double-ring graph. Two ring-based graphs are connected node to node. }
    \label{fig:topology}
\end{figure}

\begin{figure}
    \centering
    \includegraphics[width=0.48\linewidth]{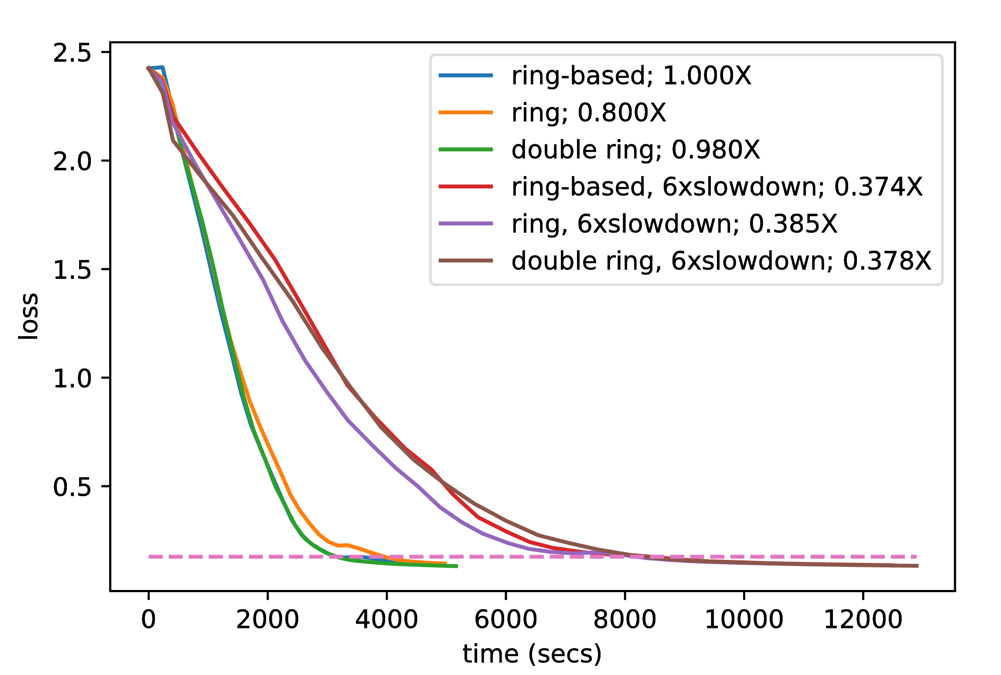}
    \includegraphics[width=0.48\linewidth]{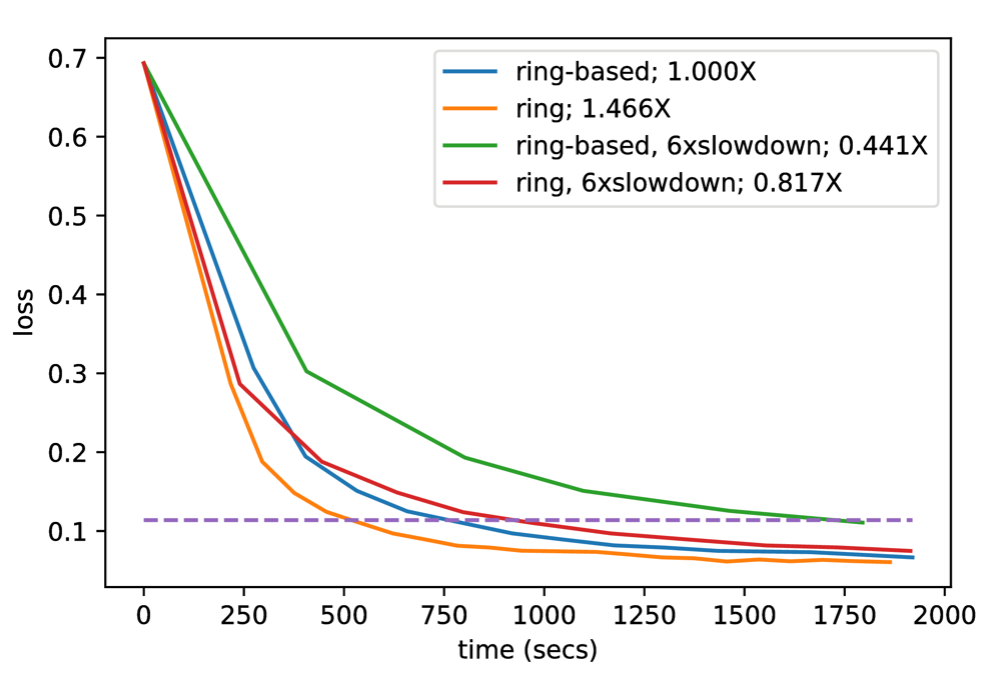} 
    \caption{Effect of heterogeneity (left: CNN; right: SVM)}
    \label{fig:2.hetero}
\end{figure}

We simulate a heterogeneous environment by randomly slowing down every worker by 6 times at a probability of $1/n$ in each iteration, where $n$ is the number of workers. 
We conduct experiments 
with and without slowdown 
on three different communication graphs (labeled as ring, ring-based and double-ring as shown in Figure \ref{fig:topology}), and the result is illustrated in Figure \ref{fig:2.hetero}. None of the graphs is immune to the slowdown.
Moreover, we see that sparser graphs suffer less to random slowdown. 

\subsubsection{Comparison to Parameter Servers}
\begin{figure}
    \centering
    \includegraphics[width=0.48\linewidth]{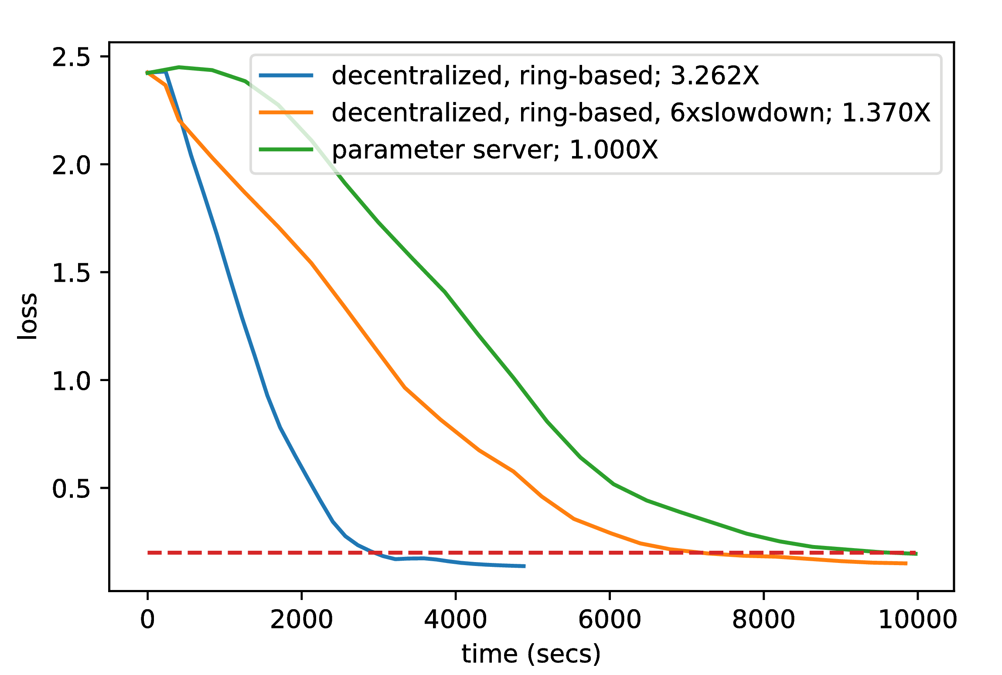}
    \includegraphics[width=0.48\linewidth]{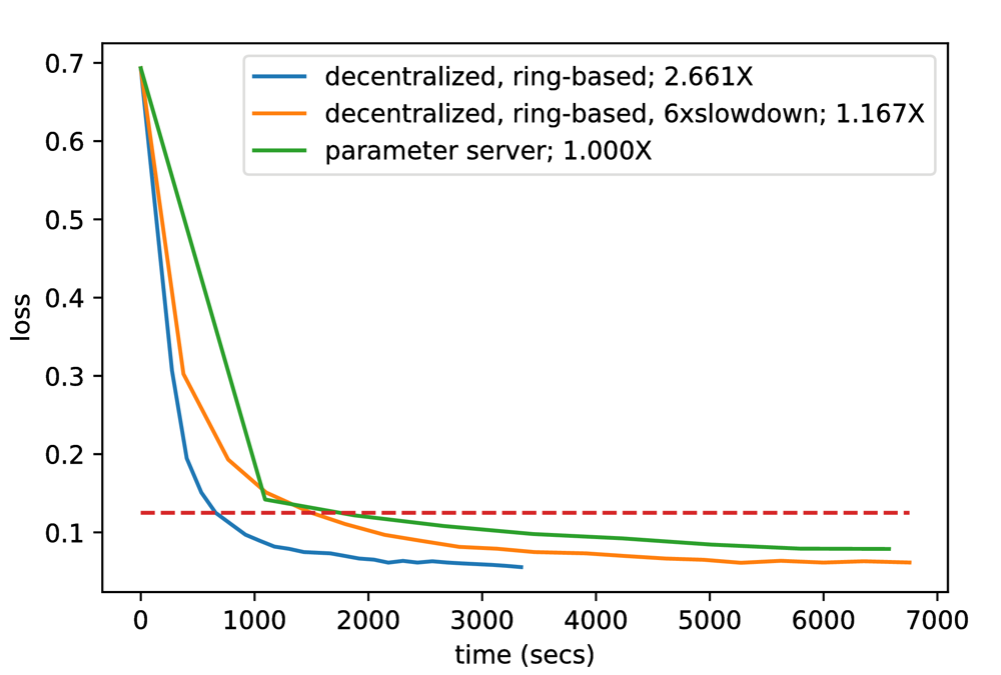}
    \caption{Decentralized vs. PS (left: CNN; right: SVM)}
    \label{fig:1.ps-vs-dc-vgg}
\end{figure}

For decentralized algorithm, training is conducted on a ring-based topology (Figure \ref{fig:topology}); for PS, we adopt BSP and use one additional machine as the parameter server. As shown in Figure \ref{fig:1.ps-vs-dc-vgg}, decentralized training in either heterogeneous or homogeneous environments converges much faster 
than homogeneous PS. Because the parameter server algorithm will inevitably be slowed down in a heterogeneous environment \cite{SIGMOD2017_het}, we do not conduct this experiment. 

\subsubsection{Effect of Backup Workers}
We design backup workers mainly for random heterogeneity, since in an environment with deterministic heterogeneity (e.g., when a worker runs much slower), the whole process will still slow down due to the token limit. 

\begin{figure}
    \centering
    \includegraphics[width=0.48\linewidth]{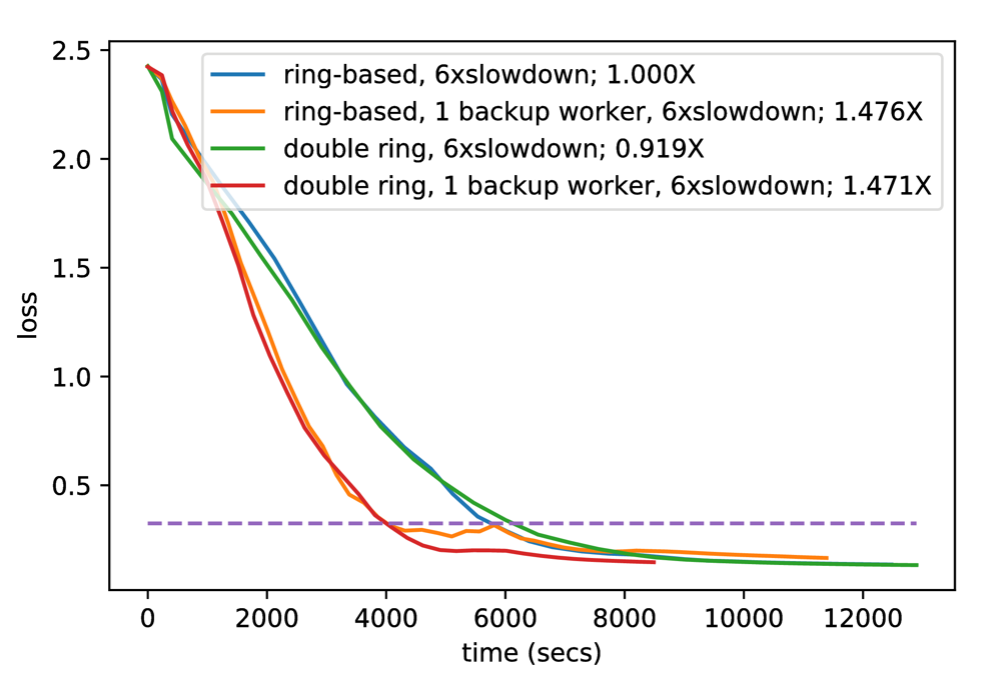}
    \includegraphics[width=0.48\linewidth]{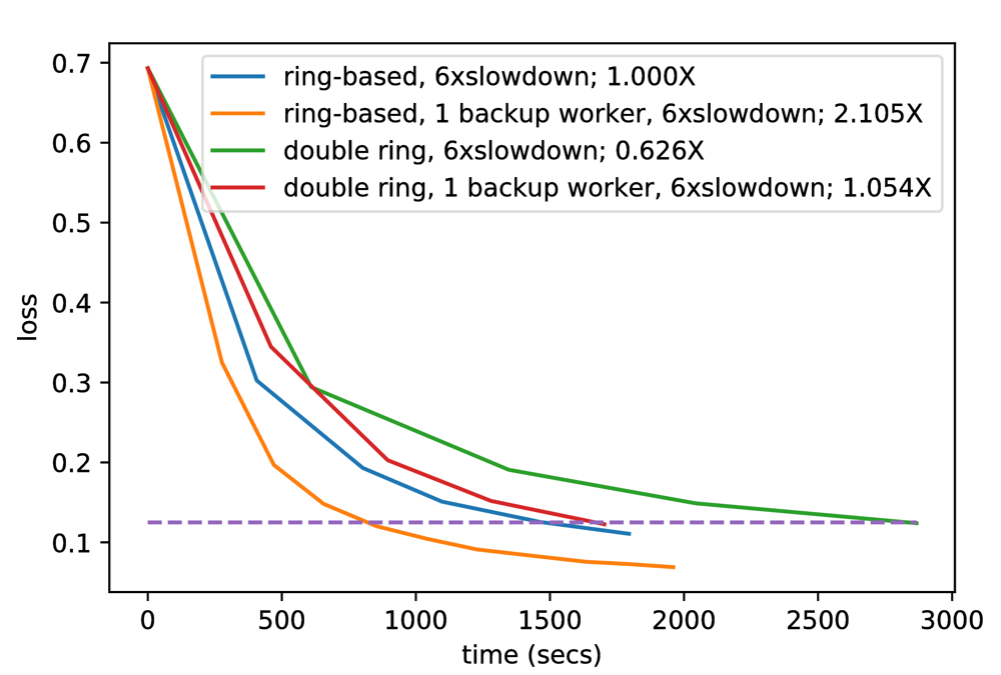}
    \caption{Effect of backup workers on decentralized training with random slowdown: loss vs time (left: CNN; right: SVM)}
    \label{fig:backup.vgg}
\end{figure}

\begin{figure}
    \centering
    \includegraphics[width=0.48\linewidth]{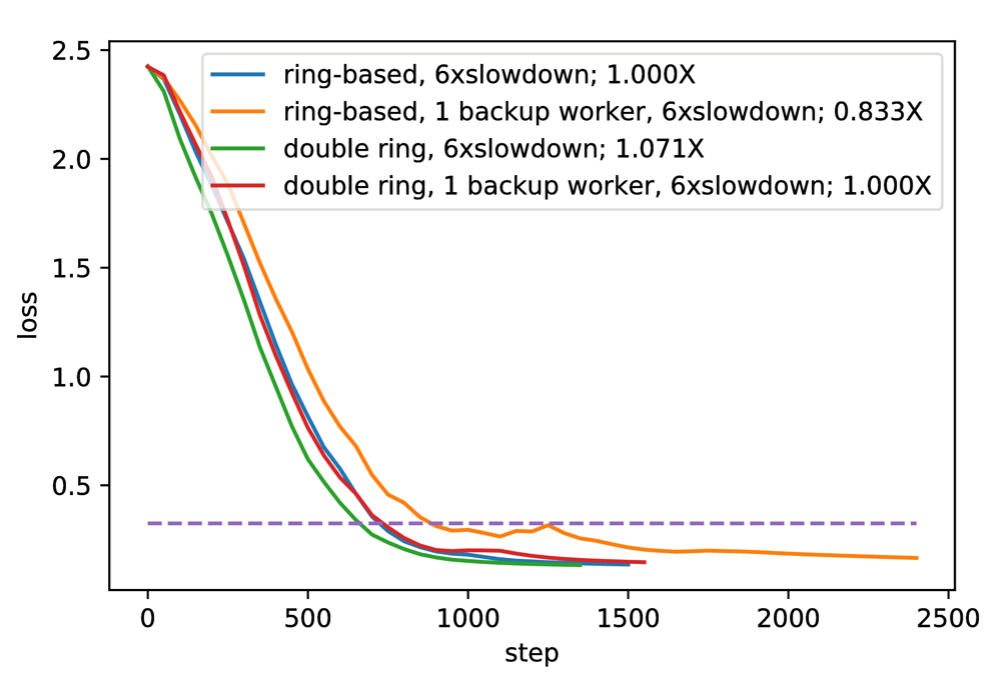}
    \includegraphics[width=0.48\linewidth]{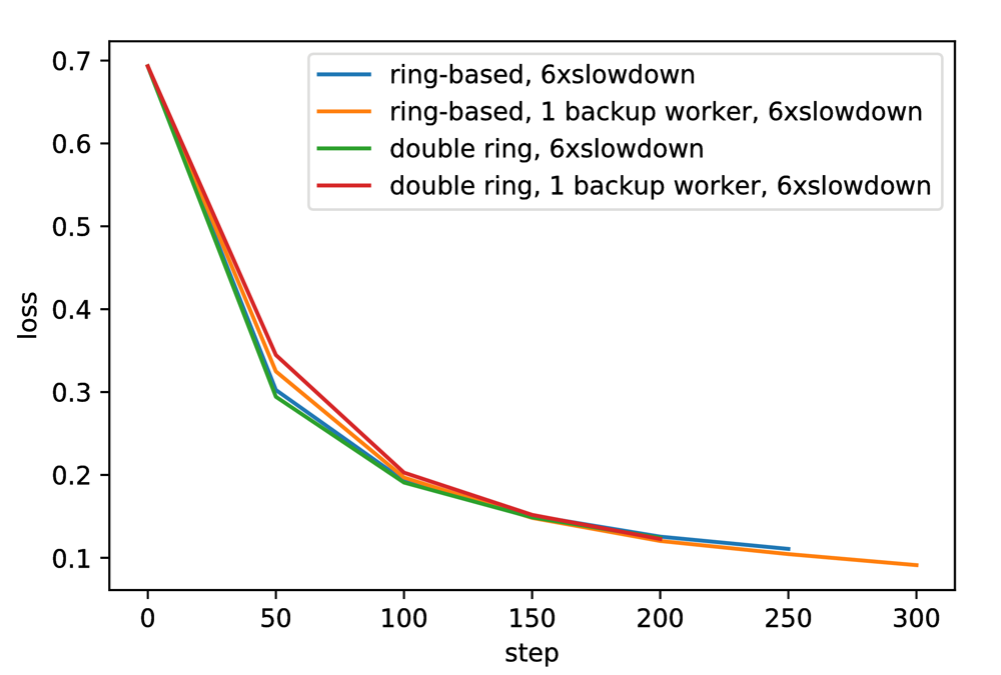}
    \caption{Effect of backup workers on decentralized training with random slowdown: loss vs steps (left: CNN; right: SVM)}
    \label{fig:backup-step.vgg}
\end{figure}

\begin{figure}
    \begin{minipage}{.45\linewidth}
    \centering
    \includegraphics[width=\linewidth]{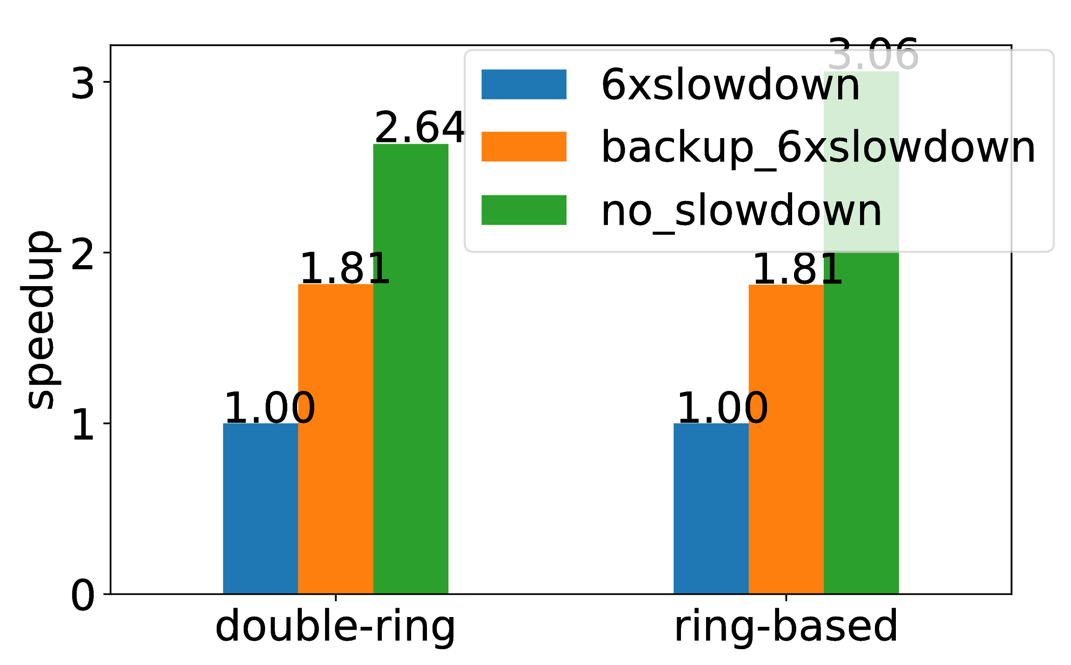}
    \caption{Effect of backup workers: iteration speed over 6Xslowdown (on CNN)}
    \label{fig:iter_speed}
    \end{minipage}
    \quad
    \begin{minipage}{.45\linewidth}
    \centering
    \includegraphics[width=\linewidth]{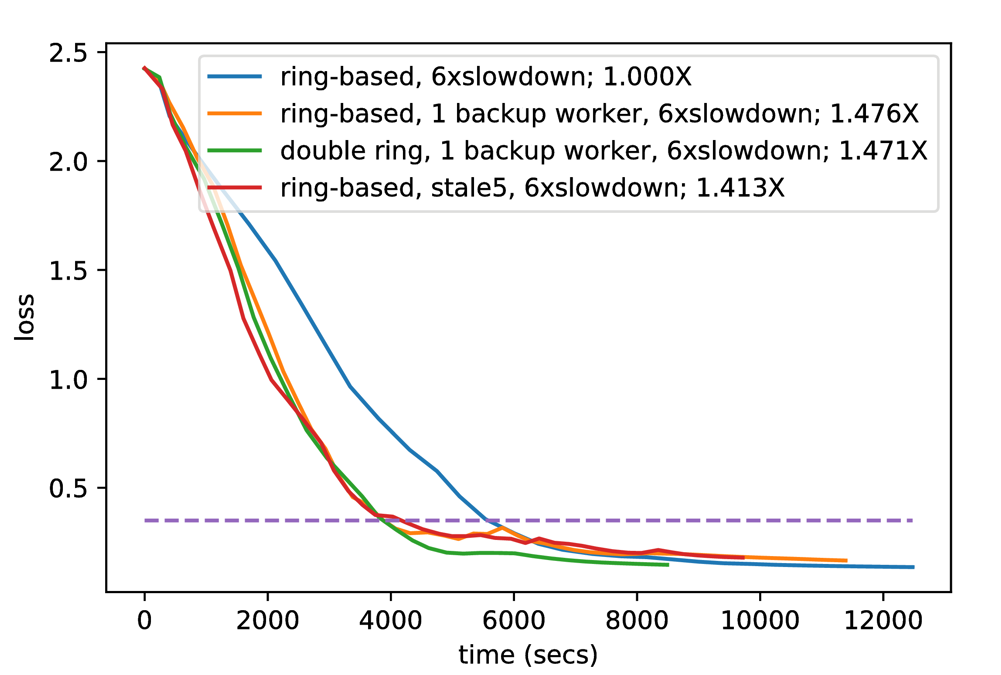}
    \caption{Effect of bounded staleness with random slowdown (on CNN)}
    \label{fig:staleness}
    \end{minipage}
\end{figure}

We test our system on 2 different communication graphs, the ring-based graph and the double-ring graph. We use one backup worker (i.e., each node can receive one less update), and the results on two graphs are similar as shown in Figure \ref{fig:backup.vgg}: training with backup workers converges faster than standard decentralized algorithms on wall-clock time. Combined with the loss curves on steps as shown in Figure \ref{fig:backup-step.vgg}, we argue that although receiving one less update hurts the
per iteration progress,
the effect is insignificant compared to the gained speedup in the per iteration execution time (a speedup of up to 1.81 is shown in Figure \ref{fig:iter_speed}).



\subsubsection{Effect of Staleness}


We conduct experiments on a ring-based graph using 6 times random slowdown and a staleness bound of 5.
As shown in \ref{fig:staleness}, the system with staleness can achieve a similar speedup to that with backup workers, and they both outperform the standard decentralized setting.

\subsubsection{Effect of Skipping Iterations}
\begin{wrapfigure}[10]{r}{0.5\columnwidth}
\vspace{-0.7cm}
\begin{center}
\includegraphics[width=0.50\columnwidth]{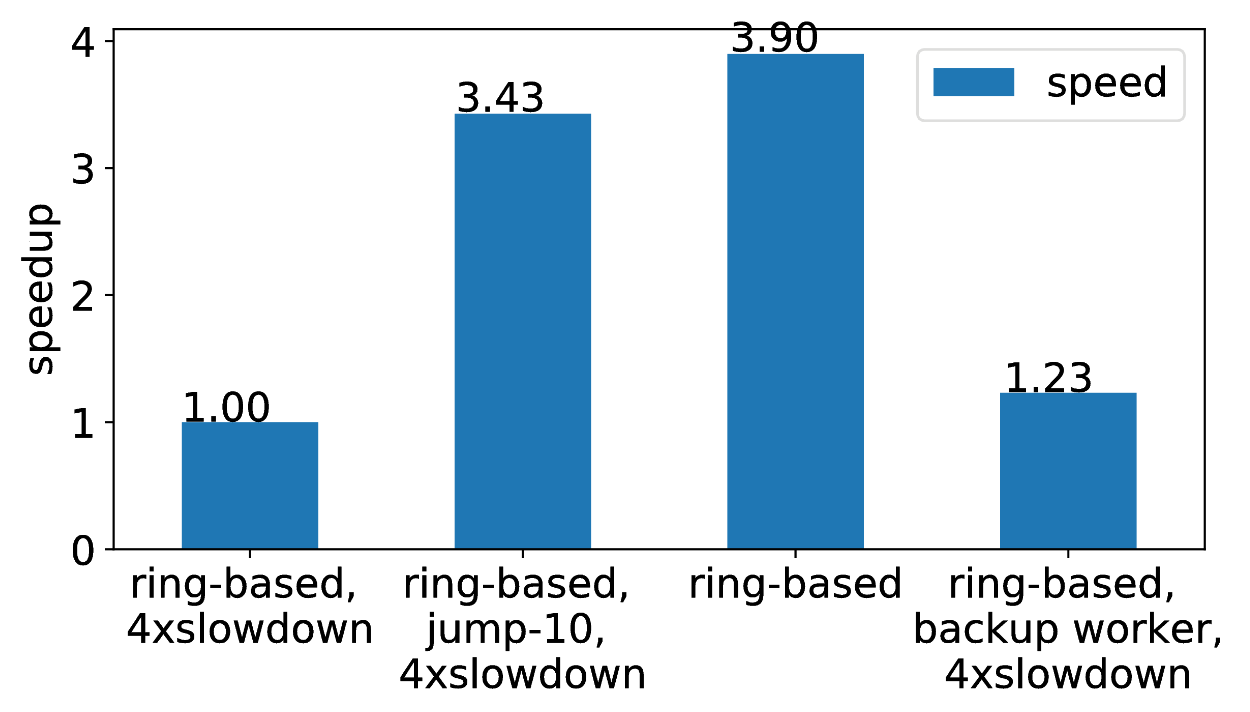}
\caption{Effect of skipping iterations: 4Xslowdown (on CNN)}
\label{fig:jump-speed}
\end{center}
\end{wrapfigure}
\begin{figure}
    \centering
    \includegraphics[width=0.48\linewidth]{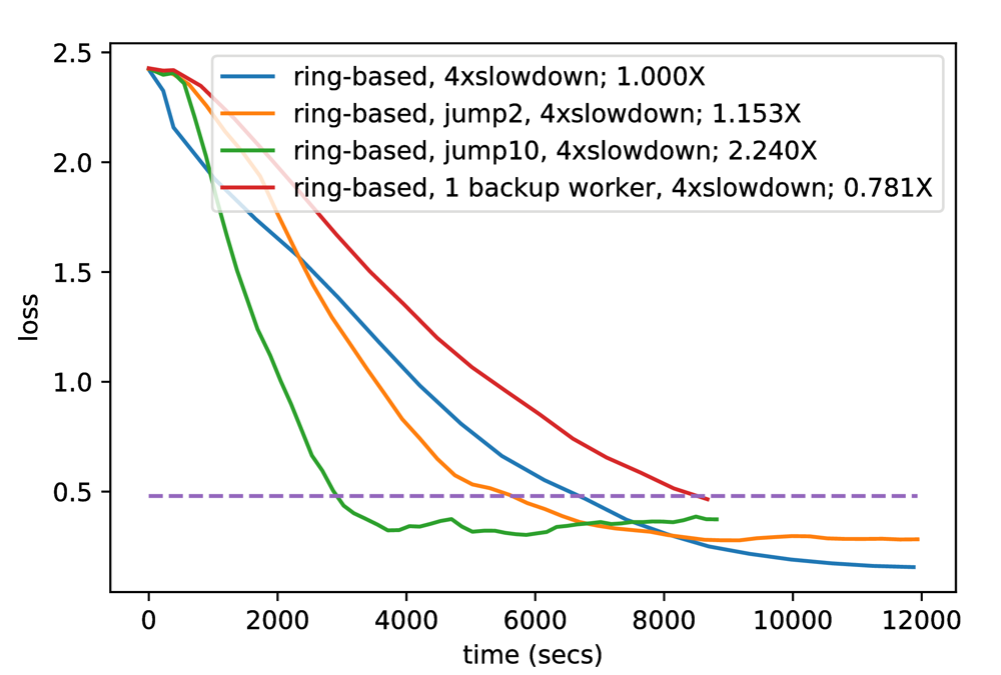}
    \includegraphics[width=0.48\linewidth]{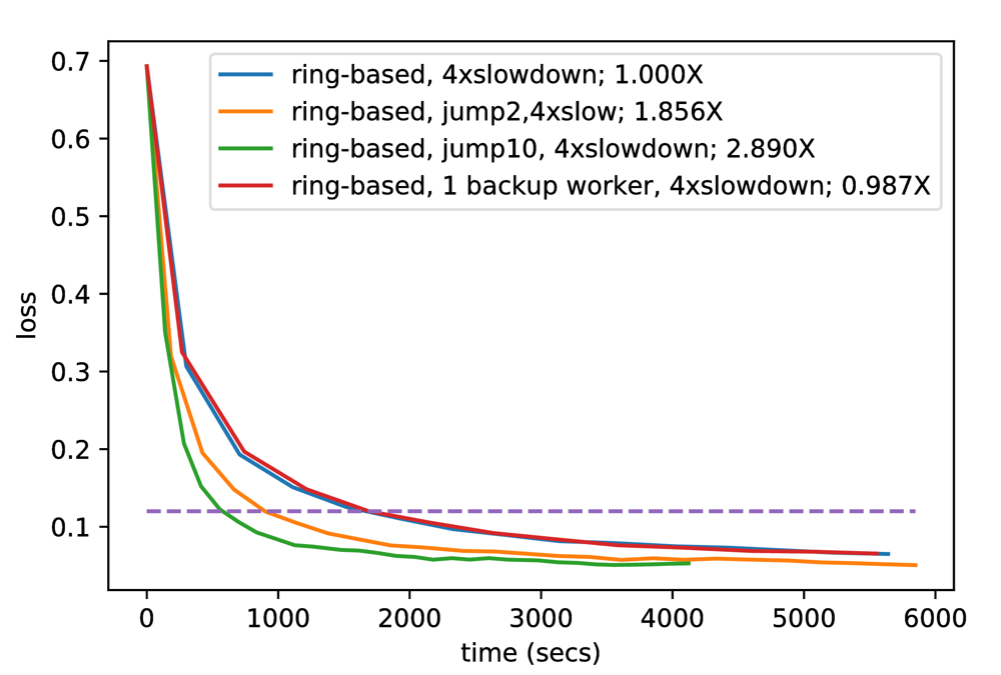}
    \caption{Effect of skipping iterations (left: CNN; right: SVM)}
    \label{fig:jump}
\end{figure}

Experiments are conducted on a ring-based graph with 16 workers, while one worker is deterministically chosen for a 4 times slowdown.
We test two settings: jumping at most 2 iterations at a time and jumping at most 10 at a time. As shown in Figure \ref{fig:jump}, skipping iterations exhibits superior performance over the simple backup worker setting, and jumping at most 10 iterations delivers the fastest convergence, with a speedup of more than 2 times over the standard decentralized system. Moreover, Figure \ref{fig:jump-speed} shows that with skipping iterations, the influence of stragglers on the duration of a iteration can be reduced a lot --- from 3.9 times slowdown to 3.90/3.43 $\approx$ 1.1 times slowdown, which contributes to the significant convergence speed gain on wall-clock time.

\label{sec:jump}

\subsubsection{Effect of graph topology}
\begin{figure}
    \centering
    \includegraphics[width=0.48\linewidth]{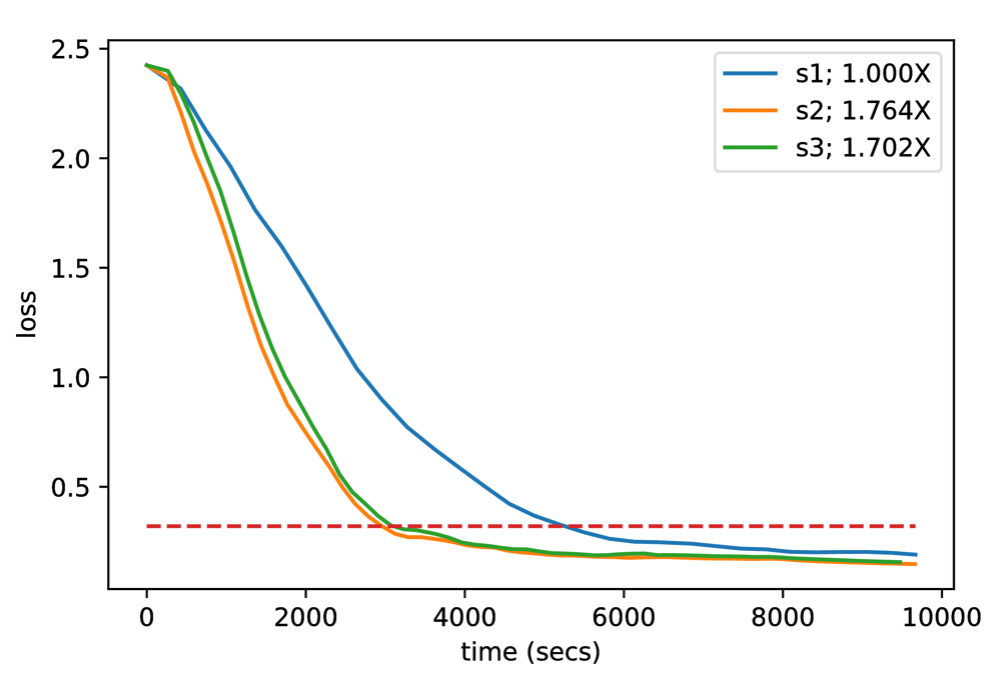}
    \includegraphics[width=0.48\linewidth]{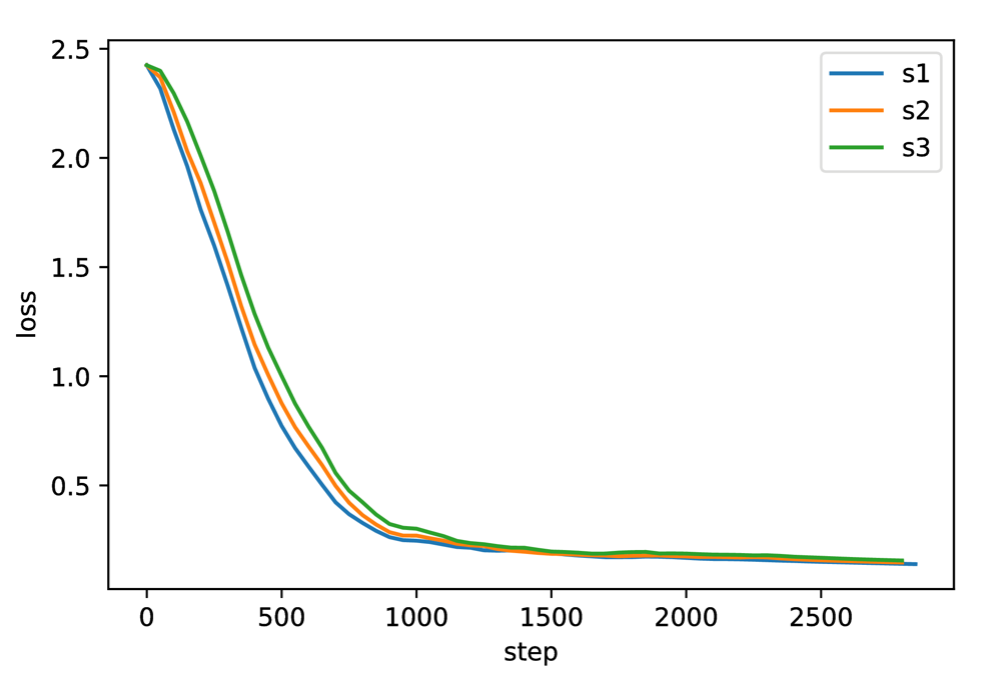}
    \caption{Comparison of three different topology settings (on CNN)}
    \label{fig:3graphs}
\end{figure}

\begin{figure}
    \centering
    \includegraphics[width=0.9\linewidth]{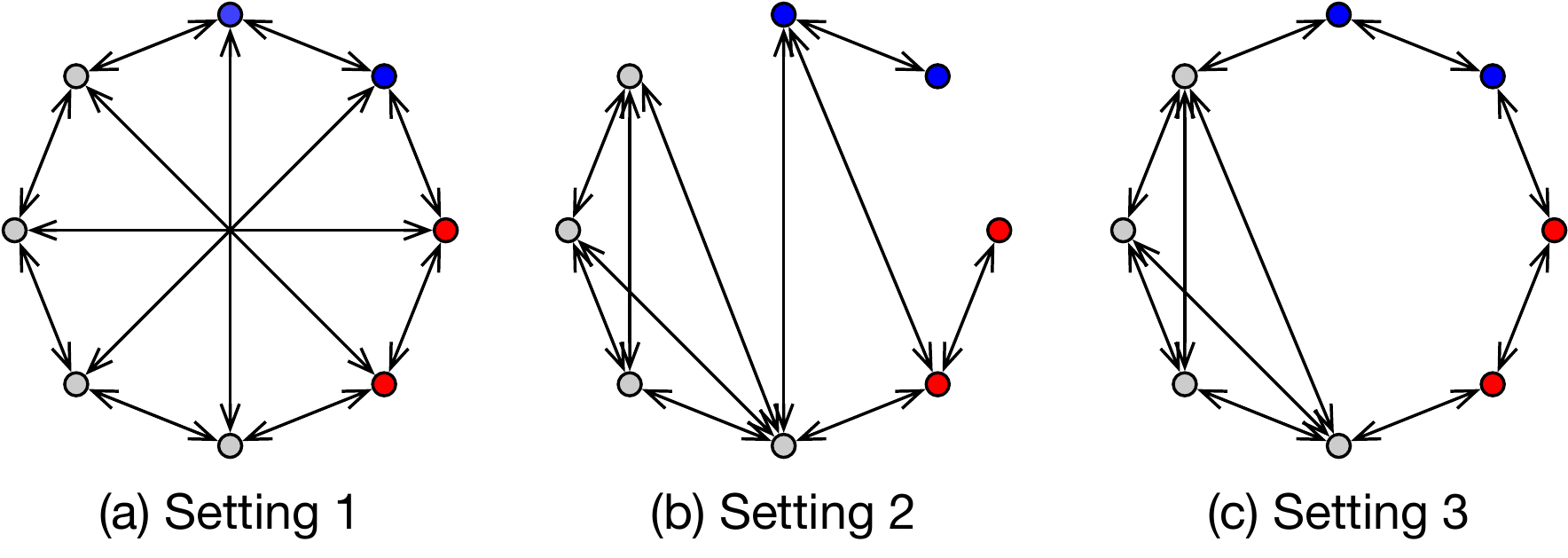}
    \caption{Three graphs tested in a heterogeneous environment (self-loops are omitted). Nodes in the same color reside in the same physical machine. Spectral gaps: (a) 0.6667, (b) 0.2682, and (c) 0.2688. }
    \label{fig:expGraph}
\end{figure}

We have compared 3 graphs in a heterogeneous setting where 8 workers are unevenly distributed over 3 machines (Figure \ref{fig:expGraph}). The baseline graph is the ring-based graph with a high spectral gap~\footnote{The spectral gap of a graph $G$ is defined as the difference between the norms of the largest 2 eigenvalues of the weighted adjacency matrix $W$, i.e. $\|\lambda_1(W)\|-\|\lambda_2(W)\|$. The bigger the spectral gap, the faster information spreads over the graph.} of 0.6667. Our proposed 2 graphs are inspired by the heterogeneous distribution of workers: an all-reduce graph is used within a physical machine, while a ring graph is used between different machines. They have much smaller spectral gaps, 0.2682 and 0.2688 respectively, but our experiments show that they perform better than the symmetric ring-based graph (Figure \ref{fig:3graphs}). In theory, the bigger the spectral gap, the fewer iterations it takes to converge \cite{ArXiv_ASAP,NIPS2017_dPSGD}. However, our experiments do not show a significant difference in the convergence rate w.r.t iterations, even when the spectral gaps are very dissimilar (Figure \ref{fig:3graphs}). Moreover, the duration of an iteration can largely vary due to the graph topology as well as the heterogeneity in the system, which has provided insights that more factors should be taken into consideration when designing the communication graph.


\section{Conclusion}
\label{sec:conclsn}
This paper proposes \projectname, a heterogeneity-aware decentralized training protocol.
Based on a unique characteristic of decentralized training that we have identified, the iteration gap, we propose a queue-based synchronization mechanism that can efficiently implement backup workers and 
bounded staleness in the decentralized setting. 
To cope with
deterministic slowdown, we propose skipping iterations 
so that the effect of slower workers is further mitigated.
We build a prototype implementation of 
\projectname on \textsc{TensorFlow}.
The experiment results on CNN and SVM
show significant speedup over 
standard decentralized training in heterogeneous settings.



\bibliographystyle{plain}
\bibliography{ref}

\end{document}